\documentclass[%
 reprint,
superscriptaddress,
longbibliography,
 amsmath,amssymb,
 aps
]{revtex4-2}
\usepackage[x11names]{xcolor}
\usepackage{graphicx,graphics,overpic}
\usepackage{dcolumn}
\usepackage{float}
\usepackage{tikz}
\usepackage[compat=1.1.0]{tikz-feynman}
\usetikzlibrary{arrows.meta,decorations.pathmorphing,calc}
\usepackage{ragged2e}
\usepackage{booktabs}
\usepackage{amsmath,amssymb,amsfonts}
\usepackage{latexsym,verbatim}
\usepackage{bm}
\usepackage{soul}

\usepackage[colorlinks=true,linkcolor=blue,citecolor=blue,urlcolor=blue]{hyperref}
\usepackage{refcount}

\usepackage{siunitx}

\usepackage{tabularx}
\newcommand{\be}{\begin{eqnarray}}
\newcommand{\ee}{\end{eqnarray}}
\newcommand{\nn}{\nonumber\\}
\newcommand{\bw}{\begin{widetext}}
\newcommand{\ew}{\end{widetext}}

\newcounter{SOM}

\begin{document}

\preprint{APS/123-QED}
\justifying
\title{Gauge phonon driven non-Fermi liquids in Dirac materials}
\author{Rutvij Gholap}
\affiliation{Department of Physics and Astronomy, The University of Manchester, Oxford Road, Manchester M13 9PL, United Kingdom}
\author{Alexey Ermakov}
\affiliation{Department of Physics and Astronomy, The University of Manchester, Oxford Road, Manchester M13 9PL, United Kingdom}
\author{Alexander Kazantsev}
\affiliation{Department of Physics and Astronomy, The University of Manchester, Oxford Road, Manchester M13 9PL, United Kingdom}
\author{Mohammad Saeed Bahramy}
\affiliation{Department of Physics and Astronomy, The University of Manchester, Oxford Road, Manchester M13 9PL, United Kingdom}
\author{Marco Polini}
\affiliation{Dipartimento di Fisica dell’Università di Pisa, Largo Bruno Pontecorvo 3, I-56127 Pisa, Italy}
\author{Alessandro Principi}
\affiliation{Department of Physics and Astronomy, The University of Manchester, Oxford Road, Manchester M13 9PL, United Kingdom}

\date{\today}

\begin{abstract}
We identify overdamped gauge phonons as a microscopic route to non-Fermi-liquid behavior in Dirac materials that does not require proximity to a quantum critical point. Coupling to electronic currents rather than densities, these modes realize a lattice analogue of electromagnetic coupling to a vector potential. We show that the low-energy behavior is governed by the orbital susceptibility $\chi_{\rm orb}$ and the dimensionless phonon damping strength, $\bar\alpha$. Strong damping, $\bar\alpha\gg1$, produces pronounced departures from Fermi-liquid theory. For diamagnetic systems, $\chi_{\rm orb}<0$, Fermi-liquid behavior survives only in a parametrically narrow infrared window before crossing over to non-Fermi-liquid scaling. For paramagnetic systems, $\chi_{\rm orb}>0$, it is absent altogether, yielding a marginal Fermi liquid at the lowest energies and non-Fermi-liquid scaling above. These results suggest that gauge phonons can provide a microscopic mechanism for anomalous metallic behavior in systems such as twisted bilayer graphene.
%We identify overdamped gauge phonons as a new microscopic route to non-Fermi-liquid behavior in Dirac materials that does not require proximity to a quantum critical point. Because gauge phonons couple to electronic currents rather than densities, they realize a lattice analogue of the coupling to an electromagnetic vector potential. We show that the low-energy behavior is controlled by the system's orbital susceptibility $\chi_{\rm orb}$ and by the (dimensionless) damping strength of transverse currents, $\bar\alpha$. Strong damping, $\bar\alpha\gg1$, drives pronounced departures from Fermi-liquid theory. When the system is diamagnetic, $\chi_{\rm orb}<0$, the Fermi-liquid regime is confined to a parametrically narrow infrared window before crossing over to non-Fermi-liquid scaling. For a paramagnetic susceptibility, $\chi_{\rm orb}>0$, the Fermi-liquid regime is entirely absent. Instead, the system is a marginal Fermi liquid and exhibits non-Fermi-liquid scaling at higher energies. Our results suggest that gauge phonons may offer a microscopic route to anomalous metallic behavior in systems such as twisted bilayer graphene.
\end{abstract}

\maketitle

%\tableofcontents

\textit{Introduction---}Landau's Fermi-liquid (FL) theory describes interacting metals in terms of long-lived quasiparticles whose zero-temperature decay rate with the quasiparticle energy $\varepsilon$ as $\Gamma\propto\varepsilon^2$ near the Fermi surface~\cite{AGD,Landau1957FL,Landau1957Osc}. 
%{\color{blue}(Marco: since this is for broad audience, shall we say what $\epsilon$ is very briefly?)} 
This scaling makes quasiparticles asymptotically sharp. Yet many correlated metals display robust non-Fermi-liquid (NFL) phenomenology, most prominently an approximately $T$-linear resistivity over broad~\cite{Bruin2013,phillips2022stranger,hartnoll2022planckian,checkelsky2024flat,hu2024quantum}.

%{\color{blue}(Marco: this is super broad research field, Planckian and strange metals. I do understand citing the first paper by Andy Mackenzie but the second one makes no sense. There are zillions of experimental papers. Long time ago I do remember that I suggested citing a review. I insist on that. Citing only one experimental paper is like commiting suicide. There is at least two modern and authoritative reviews on the topic of linear-in-$T$ resistivity. By the way, references are not abbreviated according to APS style and in modern APS style they should also have titles.)}

Identifying generic microscopic mechanisms that destroy FL behavior is therefore a central problem. Known routes to NFL behavior include coupling to gapless collective modes. The marginal-Fermi-liquid (MFL) phenomenology gives $\Gamma\propto\varepsilon$~\cite{varma1989marginal}, while quantum-critical metals described by Hertz--Millis theory~\cite{Hertz1976,Millis1993} can exhibit stronger singularities, including $\Gamma(\varepsilon)\propto\varepsilon^{2/3}$ in two dimensions. More generally, the scaling depends on the universality class, the critical fluctuations, and possible crossovers~\cite{Lohneysen2007,ChubukovReview,MetlitskiSachdev2010,Mross2010,HooleyPRL2015,HooleyPRB2018}. Spatial randomness can further stabilize strange-metal phenomenology, as shown for random Yukawa couplings in two-dimensional quantum-critical metals~\cite{Patel2023Universal,patel2024localization}. 

%{\color{blue}(Marco: are you sure that this is still the only paper by Patel and Subir on the topic?)} 

A different class of NFLs arises from overdamped transverse gauge fluctuations, which couple to currents rather than densities and need not be tied to an order-parameter transition~\cite{Polchinski1994,AltshulerIoffeMillis1994,Lee2009GaugeNFL,LeeNagaosaWen2006,HooleyPRR2025}. Itinerant electron systems confined to suitable optical cavities provide another current-coupled platform, where NFL behavior can emerge near the onset of a superradiant phase transition~\cite{rao2023non}. %{\color{blue}(Marco: again, the field has zillions of reviews and picking up only one paper does not make sense. Remove reference to  the 2020 Phys Rev Research paper by Rokaj et al.. In this context it is OK to cite only the work by Piazza.)}
More broadly, theories without well-defined quasiparticles, including SYK-inspired constructions, offer generic routes to NFL behavior~\cite{chowdhury2022sachdev}.

\begin{figure}[b]
\includegraphics[width=0.48\textwidth]{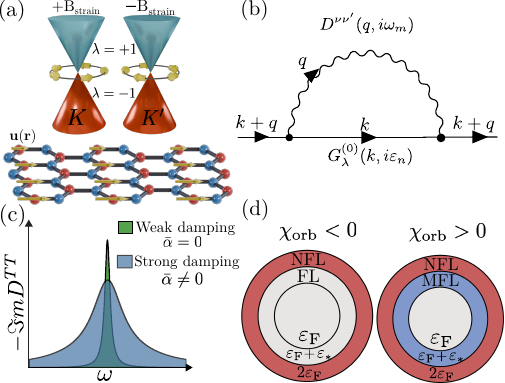}
\caption{Schematic illustration of the model and results.
(a) Lattice distortions due to a displacement field $\textbf{u}(\textbf{r})$ and the low-energy electronic structure, with strain inducing opposite pseudomagnetic fields ($\pm B_{\rm strain}$) in the $K$ and $K'$ valley. 
(b) $G_0W$ electron-phonon self-energy. Arrows denote the bare electronic Green's functions, while the wiggly line stands for the dressed phonon propagator.
(c) Representative transverse phonon spectral functions $-\mathrm{Im}\,D^{\rm{TT}}(q,\omega)$ in the weak and strong damping regimes, controlled by $\bar{\alpha}$. 
(d) The resulting crossover between FL/NFL and MFL/NFL regimes, for $\chi_{\rm orb}<0$ and $\chi_{\rm orb}>0$, respectively.}
\label{fig:1}
\end{figure}

In this Letter, we show that \emph{gauge phonons}~\cite{suzuura2002phonons,manes2007symmetry,Vozmediano2010,Sohier2014PRB_GaugePhononResistivity}
%{\color{blue}(Marco: are you sure this is the first paper on the topic? Otherwise why picking this particular one up? I remember this is where we started from long time ago but this does not mean that it is the first introducing gauge phonons.)} 
provide a solid-state route to current-mediated NFL physics, as schematically shown in Fig.~\ref{fig:1}. Such phonons couple to electronic currents rather than densities and arise naturally in crystals with non-orthogonal lattice vectors. 
%Related pseudogauge-field constructions have also been analyzed in suitably engineered square-lattice settings~\cite{sun2023square}. 
%
For the sake of concreteness, we focus on Dirac materials, where lattice deformations generate effective valley-dependent gauge fields~\cite{Vozmediano2010,Guinea2010NatPhys}. We discuss the case of systems with reduced Fermi velocity, potentially relevant to magic-angle twisted bilayer graphene (MATBG) \cite{bistritzer2011moire,Cao2018Insulator,Cao2018Superconductivity}%{\color{blue}(Marco: mentioning MATBG without citing Bistritzer-MacDonald 2011 model + the two 2018 papers by Pablo is offensive and scientific immoral.)}
, where linear-in-temperature resistivity and strange metallic behavior has been reported~\cite{yuancao,polshyn2019large}.
%{\color{blue}(Marco: are you sure Andrea Young's paper spotted before Pablo the linear-in-$T$ behavior and made contact to Planckian behavior? Here I would feel much more comfortable citing only Pablo's paper and follow up works. Be extremely carefuly here for reasons I do not want to mention.)}%
However, the existence of a gauge coupling to electronic currents does not rely on hexagonal symmetry, it is generic to non-square or distorted square lattices. Therefore, the microscopic mechanism discussed here, which requires only phonons rather than additional critical bosonic modes, may be applicable to a broad range of quantum materials.

We compute the electronic self-energy from a transverse phonon propagator dressed by the electron-phonon interaction. The low-energy physics is controlled by the orbital magnetic susceptibility $\chi_{\rm orb}$ and a dimensionless damping parameter $\bar\alpha$. For $\bar\alpha\gg1$, gauge phonons become strongly damped and produce pronounced NFL effects. If $\chi_{\rm orb}<0$, the system remains FL-like only below an extremely small scale and crosses over at $\varepsilon_\ast$ to NFL behavior. If $\chi_{\rm orb}>0$, phonon softening eliminates the conventional FL infrared regime: a MFL regime appears instead, before crossing over to NFL scaling at higher energies. We stress that this MFL behavior occurs not only in charge-neutral Dirac systems~\cite{das2007many,barlas2009non},
%, where related interaction-driven MFL or NFL behavior has been discussed, 
but also at finite doping, where 
%the noninteracting system would be 
electron-electron interactions alone would produce
an ordinary FL \cite{tsvelik2003quantum}.
%{\color{blue}(Marco: are you referring to the fact that gauge-field-mediated effective electron-electron interactions remain unscreened even in high-density metals? If yes, I would cite the nice book by Tsvelik here where there is an entire chapter on this -- Electrodynamics of metals.)}
Our results have therefore potential implications for the strange metallicity observed in MATBG~\cite{polshyn2019large,yuancao}. However, fully establishing MATBG as a potential platform for NFL physics mediated by gauge phonons would require a detailed microscopic analysis, which is beyond the scope of this work.

\textit{Model}---We consider electrons described by a generic non-interacting Hamiltonian
${\cal H}_e=\sum_{{\bm k},\lambda}\xi_{{\bm k},\lambda}\,
c^\dagger_{{\bm k},\lambda}c_{{\bm k},\lambda}$,
where $\xi_{{\bm k},\lambda}=\varepsilon_{{\bm k},\lambda}-\mu$ is the band energy measured from the chemical potential $\mu$, and $\lambda=+1$ ($\lambda=-1$) labels the conduction (valence) band. For concreteness, we approximate the band dispersion by that of massless Dirac fermions,
$\varepsilon_{{\bm k},\lambda}=\lambda v_{\rm F}k$,
where $v_{\rm F}$ is the Fermi velocity. Hereafter, we set $\hbar = 1$.
The gauge-phonon dynamics is described by the Hamiltonian
${\cal H}_{\rm ph}=\sum_{{\bm q},\nu}\omega_{{\bm q},\nu}\,
a^\dagger_{{\bm q},\nu}a_{{\bm q},\nu}$,
where $\omega_{{\bm q},\nu}=c_{{\rm ph},\nu}q$ is the acoustic-phonon
dispersion, $c_{{\rm ph},\nu}$ is the sound velocity, and
$\nu={\rm L},{\rm T}$ distinguishes longitudinal and transverse modes. Lattice distortions modify the electronic hopping amplitudes and generate
both scalar and vector electron-phonon~\cite{Vozmediano2010,Vozmediano2010PhysRep} %{\color{blue}(Marco: citing again reviews here by Maria, Paco, and Misha)}
. The corresponding low-energy Hamiltonian in one of the two valleys can be written as
\begin{eqnarray}
\label{eq:e_ph_ham}
{\cal H}_{\rm e\text{-}ph}
=
\sum_{{\bm k},{\bm q},\alpha,\beta}
c^\dagger_{{\bm k}-{\bm q},\alpha}
\Big({v_{\rm F}\bm A}_{\bm q}\!\cdot\!{\bm {\sigma}}
+V_{\bm q}\openone\Big)_{\alpha\beta}
c_{{\bm k},\beta}
~,
\end{eqnarray}
where $V_{\bm q}$ is the scalar deformation-potential coupling,
${\bm A}_{\bm q}$ is the strain-induced (valley-dependent) gauge potential, and $\bm \sigma = (\sigma_{x}, \sigma_{y})$ are Pauli matrices acting in sublattice space. 
% We note that both phonon polarizations contribute to the longitudinal and transverse components of the gauge potential, $A_{{\bm q},L}$ and $A_{{\bm q},T}$, respectively. 
In the following we focus on the latter, since the scalar potential 
% $V_{\bm q}$ 
is screened by electron-electron interactions \cite{gibertini2012puddles}.%{\color{blue}(Marco: I would cite some work here. You can look up old papers by Gibertini and myself: https://doi.org/10.1103/PhysRevB.85.201405 and tell the reader to look into references there, where we provided a full review of precisely this fact, fights above precise values of the deformation potential, etc etc etc.)}
% , whereas the gauge potential is not, . 
Further details of ${\cal H}_{\rm e\text{-}ph}$ are provided in \footnote{See the Supplemental Online Material.}\setcounter{SOM}{\value{footnote}}.

\textit{Self-energy}---We consider the $G_0W$ electron-phonon self-energy
$\Sigma_\lambda({\bm k},i\varepsilon_n)$ shown in Fig.~\ref{fig:1}(b):
\begin{align}
&\Sigma_\lambda({\bm k},i\varepsilon_n)
=
-\beta^{-1}
\sum_{\omega_m}
\sum_{\nu,\nu',\lambda'}
\int\frac{d^2{\bm q}}{(2\pi)^2}
D^{\nu,\nu'}({\bm q},i\omega_m)
\nonumber \\
&\times
G^{(0)}_{\lambda'}({\bm k}+{\bm q},i\varepsilon_n+i\omega_m)
j^{\nu}_{{\bm k},\lambda;{\bm k}+{\bm q},\lambda'}
j^{\nu'}_{{\bm k}+{\bm q},\lambda';{\bm k},\lambda}.
\label{eq:SE_def}
\end{align}
Here $\varepsilon_n=(2n+1)\pi/\beta$ and $\omega_m=2m\pi/\beta$
are fermionic and bosonic Matsubara frequencies, respectively,
$D^{\nu,\nu'}({\bm q},i\omega_m)$ is the phonon propagator dressed by the electron-phonon interaction
and
$
G^{(0)}_{\lambda}({\bm k},i\varepsilon_n)
=
(i\varepsilon_n-\xi_{{\bm k},\lambda})^{-1}
$
is the bare electron Green's function. 
Each valley contributes independently to the self-energy. In multivalley Dirac systems with well-separated valleys,
inter-valley scattering requires phonons with a large momentum and is highly ineffective~\cite{borysenko2010first,gunst2016first}. 
%{\color{blue}(Marco: we should be talking about Dira materials in general, no? Not only graphene.)}
Since we neglect inter-valley mixing and excitations, the two contributions simply add up as if currents were coupled to an electromagnetic vector potential.

After analytical continuation $i\varepsilon_n\to\varepsilon+i0^+$,
%the imaginary part of the retarded self-energy can be written as
%\begin{align}
%&\mathrm{Im}\Sigma_\lambda^{(R)}({\bm k},\varepsilon)
%=
%-\int_{-\infty}^{\infty}\frac{d\omega}{\pi}
%\big[n_{\rm B}(\omega)+n_{\rm F}(\omega+\varepsilon)\big]
%\nonumber \\
%&\times
%\sum_{\nu,\nu',\lambda'}
%\int\frac{d^2{\bm q}}{(2\pi)^2}
%\mathrm{Im} D^{\nu,\nu'}({\bm q},\omega)
%\mathrm{Im} G^{(0)}_{\lambda'}({\bm k}+{\bm q},\varepsilon+\omega)
%\nonumber \\
%&\times
%j^{\nu}_{{\bm k},\lambda;{\bm k}+{\bm q},\lambda'}
%j^{\nu'}_{{\bm k}+{\bm q},\lambda';{\bm k},\lambda}.
%\label{eq:sigma_analit_3}
%\end{align}
%
%We now 
we can focus on the physics at the Fermi surface.
For $T\to0$ the chemical potential
coincides with the Fermi energy of the non-interacting system,
$\mu=\varepsilon_{\rm F}=v_{\rm F}k_{\rm F}$. Here $k_{\rm F}$ is the Fermi wave vector. 
%the factor
%$n_{\rm B}(\omega)+n_{\rm F}(\omega+\varepsilon)$ restricts 
The frequency integral 
is restricted
to the interval $0<\omega < |\varepsilon|\ll \varepsilon_{\rm F}$, while
intermediate fermionic states are pinned at the Fermi surface.
% The matrix elements of the current operator, therefore, simplify to
% \begin{align}
% j^{\rm L}_{{\bm k},\lambda;{\bm k}',\lambda'}
% &=
% v_{\rm F}\frac{i(\lambda'-\lambda)}{2}
% e^{i(\varphi_{{\bm k}'}-\varphi_{\bm k})/2},
% \label{eq:sigma_LT_FS_def_L}
% \\
% j^{\rm T}_{{\bm k},\lambda;{\bm k}',\lambda'}
% &=
% v_{\rm F}\frac{\lambda'+\lambda}{2}
% e^{i(\varphi_{{\bm k}'}-\varphi_{\bm k})/2}.
% \label{eq:sigma_LT_FS_def_T}
% \end{align}
% 
% Since 
%At the Fermi surface, 
% $\lambda'=\lambda$, 
%he longitudinal matrix
%element vanishes and 
In this limit, only the transverse channel $\nu=\nu'={\rm T}$ 
%{\color{blue}(Marco: you have to decide whether it is going to be ${\rm T}$ or $T$ everywhere, in the propagator below and Fig.~1(c). Be consistent. I of course vote for ${\rm T}$.)}
contributes to the self-energy. 
% The intermediate steps of this derivation 
More details are given in~\footnotemark[\value{SOM}].
%Appendix~\ref{app:self_energy}.

%\vspace{2mm}
%\begin{figure}
%\centering
%\resizebox{\columnwidth}{!}{%
%\begin{tikzpicture}[baseline=(current bounding box.center)]
%\begin{feynman}
%
%% vertices
%\vertex (a) at (0,0);
%\vertex [dot] (b) at (2.0,0){};
%\vertex [dot] (c) at (6.0,0){};
%\vertex (d) at (8.4,0);
%
%% propagators
%\diagram*{
%(a) -- [fermion] (b)
%    -- [fermion] (c)
%    -- [fermion] (d),
%
%(b) -- [boson, half left,  looseness=1.5, line width=0.7pt,
%        edge label={$D^{\nu,\nu'}(q,i\omega_m)$}] (c),
%};
%
%% manual labels
%\node[below=7pt, font=\footnotesize] at ($(a)!0.5!(b)$)
%{$G^{(0)}_{\lambda'}(k+q,i\varepsilon_n+i\omega_m)$};
%
%\node[below=7pt, font=\footnotesize] at ($(b)!0.5!(c)$)
%{$G^{(0)}_{\lambda}(k,i\varepsilon_n)$};
%
%\node[below=7pt, font=\footnotesize] at ($(c)!0.5!(d)$)
%{$G^{(0)}_{\lambda'}(k+q,i\varepsilon_n+i\omega_m)$};
%
%\end{feynman}
%\end{tikzpicture}
%}
%\caption{The $G_0W$ electron-phonon self-energy $\Sigma_\lambda(\bm{k}, i\varepsilon_n)$. The arrows denote bare electronic Green's functions, while the wiggly line stands for the phonon propagator dressed by electron-phonon interactions.}
%\label{fig:fig2}
%\end{figure}

 \textit{Dressed phonon propagator}---The non-interacting phonon propagator reads
\begin{equation}
D^{(0),\rm{TT}}({\bm q},\omega)
=
\sum_{\nu}
g_\nu^2(q)\,
% \nonumber\\
% &\times 
\big[ M_{{\rm T},\nu}(\hat{\bm q})\big]^2
\frac{2\omega_{{\bm q},\nu}}
{(\omega+i0^+)^2-\omega_{{\bm q},\nu}^2},
\label{eq:D0_TT_ret}
\end{equation}
where $g_\nu(q)=(g_2/v_{\rm F})\sqrt{q^2/(2\rho_m\omega_{{\bm q},\nu})}$. Here, $g_2$ is the electron-phonon coupling constant and $\rho_m$ is the crystal mass density. Note that 
% the transverse gauge field is a linear combination of longitudinal and transverse phonon modes. As a result,
both longitudinal and transverse phonon modes contribute to the transverse propagator, with the matrix $ M_{{\rm T},\nu}(\hat{\bm q})$ encoding this mixing.
For the purpose of scaling analysis, in what follows we 
% approximate the longitudinal and
% transverse acoustic 
assume both branches to have the same frequency, {\it i.e.}
% by a single effective mode of frequency
$\omega_{\bm q} \approx \omega_{{\bm q},{\rm L}}\approx \omega_{{\bm q},{\rm T}}$.
This allows us to drop all mode indices $\nu = \rm{T, L}$. %{\color{blue}(Marco: you have to decide whether it is going to be ${\rm T}$ or $T$ everywhere. Be consistent. I of course vote for ${\rm T}$.)} in the following equations.
% Note that s
Such simplification only affects overall prefactors and does not modify the
scaling exponents discussed below. 

The phonon propagator is dressed by electron-phonon interactions. Within the large-$N$ approximation, the dressed propagator is obtained via a random-phase-approximation (RPA) resummation. 
% Unlike the more familiar deformation-potential coupling, 
Because Eq.~\eqref{eq:e_ph_ham} couples the gauge potential to electronic currents,
% . As a result, 
%the polarization entering the Dyson equation 
the self-energy of phonons 
is given by the transverse current-current response $\chi^{(\rm{T})}({\bm q},\omega)$. 
%The dressed propagator therefore satisfies
%% \begin{equation}
%$
%\big[D^{TT}({\bm q},\omega)\big]^{-1}
%=
%\big[D^{(0),TT}({\bm q},\omega)\big]^{-1}
%-
%\chi^{(T)}({\bm q},\omega).
%$
%% \label{eq:Dyson_TT_inv}
%% \end{equation}
For a non-interacting electron liquid and 
% in the regime
for
$\omega\ll v_{\rm F}q\ll \varepsilon_{\rm F}$, the transverse
current-current response admits the low-energy expansion
\begin{align}
\chi^{(\rm{T})}(q,\omega)
\simeq
-\frac{4\pi \chi_{\rm orb}}{e^2\mu_0} q^2
-
i\,\alpha\frac{\omega}{v_{\rm F}q},
\label{eq:chiT_expand}
\end{align}
%
%{\color{blue}(Marco: you dropped ${\bm q}$ and, again, check whether you want ${\rm T}$ or $T$. I vote for the first, everywhere in the paper.)}
where $e$ is the electron charge and  $\mu_0$ is the magnetic permeability of vacuum (in SI units). 
The coefficient of the imaginary term, $\alpha$, describes Landau damping of transverse currents due to particle-hole
excitations, while $\chi_{\rm orb}$ is the orbital  magnetic susceptibility~\cite{giuliani2005quantum}. 
%Note that the conventional definition of the orbital magnetic susceptibility~\cite{giuliani2005quantum}, which we will later name $\chi_m$ to avoid confusion, is obtained by multiplying $\chi_{\rm orb}$ by $e^2 \mu_0/4\pi$, where $e$ is the electron charge and  $\mu_0$ is the magnetic permeability of vacuum (in SI units). 
Using Eq.~\eqref{eq:chiT_expand}, the dressed phonon propagator takes the
form
\begin{align}
D^{\rm{TT}}({\bm q},\omega)
=
\frac{2\omega_{\bm q} g^2_{\rm{T}}(q)}
{\omega^2-\omega_{\bm q}^2 + \bar\chi_{\rm orb} v_{\rm F}^2 q^4/k_{\rm F}^2+i\bar\alpha v_{\rm F} q\omega}~.
\label{eq:D_TT_final}
\end{align}
%with, $g(q) = (g_2/v_{\rm F}) \times \sqrt{q^2/(2 \rho_m { \omega}_{{\bm q},}})$. 

\begin{figure*}[t]
    \centering
    \includegraphics[width=0.98\textwidth]{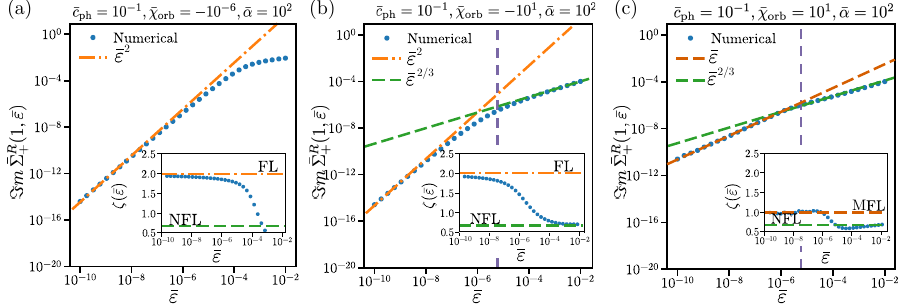}
    \caption{\label{fig:2}
    The imaginary part of the retarded self-energy as a function of the dimensionless energy. 
    Insets: The logarithmic derivative. 
    In all plots, $\bar\alpha=10^{2}$ and ${\bar c}_{\rm ph}=10^{-1}$.
    (a) For $\bar\chi_{\rm orb}=-10^{-6}$ the self-energy shows predominantly a FL behavior ($\bar{\varepsilon}_{\ast} =10^{-2}$). 
    (b) For $\bar\chi_{\rm orb}=-10^{1}$ a crossover from FL to NFL behavior occurs at $\bar{\varepsilon}_{\ast} \simeq10^{-5}$ (purple dashed vertical line). The FL behavior is thus restricted to an extremely narrow energy interval.
    (c) For $\bar\chi_{\rm orb}=10^{1}$ the FL regime is entirely absent. The crossover between MFL and NFL occurs at the same energy as panel~(b).
    }
\end{figure*}

Here we introduced the dimensionless 
orbital susceptibility and damping parameter, $\bar\chi_{\rm orb}=4 \pi g_2^2 k_{\rm F}^2 \chi_{\rm orb}/(e^2 \mu_0 v_{\rm F}^4\rho_m)$ and $\bar\alpha=(g_2^2\alpha)/(v_{\rm F}^4\rho_m)$, respectively. The sign of the orbital susceptibility determines whether the real part of the phonon
dispersion stiffens (diamagnetic $\bar\chi_{\rm orb}<0$) or softens (paramagnetic $\bar\chi_{\rm orb}>0$), while $\bar\alpha>0$
ensures that poles lie in the lower half of the
complex-frequency plane. %\color{red}(do we just hope nobody notices?)

To analyze the scaling with quasiparticle energy $\varepsilon$, it is convenient to introduce dimensionless variables in the
self-energy.
% using the Fermi wave vector
%$k_{\rm F}$. 
To this end, we define the  quantities
$\bar q=q/k_{\rm F}$,
$\bar\omega=\omega/(v_{\rm F}k_{\rm F})$,
$\bar\varepsilon=\varepsilon/(v_{\rm F}k_{\rm F})$, and ${\bar \Sigma}_+^{(R)}(1, \bar{\varepsilon}) = \Sigma_+^{(R)}(k_{\rm F}, \varepsilon)/(v_{\rm F}k_{\rm F})$, and ${\bar c}_{\rm ph} = c_{\rm ph}/v_{\rm F}$.
% , and rescale the orbital susceptibility and damping parameter as $\bar\chi_{\rm orb}\to \bar\chi_{\rm orb} k_{\rm F}^2/v_{\rm F}^{2}$ and $\bar\alpha\to \bar\alpha/v_{\rm F}$. 
The dimensionless self-energy thus becomes,
\begin{equation}
\mathrm{Im} {\bar \Sigma}_+^{(R)}(1, \bar{\varepsilon})
=
\int_{0}^{\bar{\varepsilon}}\frac{d\bar{\omega}}{\pi}
\int_{{\bar q}_0}^{{\bar q}_1}
\frac{d\bar{q}}{\pi}
%\nonumber\\
%&\times
%\frac{\bar{q} |\bar{\varepsilon}+\bar{\omega}+1|}
%{\sqrt{(2\bar{q})^2-\big[(\bar{\varepsilon}+\bar{\omega}+1)^2-(1+\bar{q}^2)\big]^2}}
%\nonumber\\
%&\times
\frac{\bar\alpha\bar{q}^3\bar{\omega} \Xi({\bar q},{\bar\omega},{\bar\epsilon})}
{(\bar{\omega}^2-{\bar \omega}_{{\rm ph},q}^2)^2
+(\bar\alpha\bar{q}\bar{\omega})^2},
\label{eq:sigma_analit_final_dimmensionless}
\end{equation}
where ${\bar q}_0 = \bar{\varepsilon}+\bar{\omega}$, ${\bar q}_1 = 2+(\bar{\varepsilon}+\bar{\omega})$ and ${\bar \omega}_{{\rm ph},q}^2 = {\bar c}^2_{\rm ph}\bar{q}^2- \bar\chi_{\rm orb}\bar{q}^4$. The function $\Xi({\bar q},{\bar\omega},{\bar\epsilon})$ does not contribute to the scaling. Its expression 
%and the
%steps leading to Eq.~\eqref{eq:sigma_analit_final_dimmensionless}
%are 
is given 
% in~\footnotemark[\value{SOM}] 
in the end matter.

\textit{Results}---Figure~\ref{fig:2} shows the numerical evaluation of Eq.~\eqref{eq:sigma_analit_final_dimmensionless} in the regime of strong damping, $\bar\alpha \gg 1$. For a diamagnetic ${\bar \chi}_{\rm orb} < 0$, at low-energies the system remains a FL, {\it i.e.} $\mathrm{Im} {\bar \Sigma}_+^{(R)}(1, \bar{\varepsilon}) \propto {\bar \varepsilon}^2$. 
%The self-energy deviates
%from the standard Fermi-liquid form 
Above the characteristic crossover scale~\footnotemark[\value{SOM}]
$\bar\varepsilon_* \sim {\bar c}_{\rm ph}^3/(\bar\alpha \sqrt{|{\bar\chi}_{\rm orb}|})$
%%
%\begin{equation} \label{eq:crossover_scale}
%\bar\varepsilon_* \sim \frac{{\bar c}_{\rm ph}^3}{\bar\alpha \sqrt{|{\bar\chi}_{\rm orb}|}}~.
%\end{equation}
%%
%Above $\bar\varepsilon_{\ast}$, 
we find $\mathrm{Im} {\bar \Sigma}_+^{(R)}(1, \bar{\varepsilon}) \propto {\bar \varepsilon}^{2/3}$. 
Figures~\ref{fig:2}(a) and~(b) show this crossover for different values of $\bar\chi_{\rm orb}$ and
$\bar\alpha$. The location of $\bar\varepsilon_*$ shifts accordingly. The corresponding logarithmic derivative $\zeta(\bar\varepsilon) \equiv d\ln \mathrm{Im}{\bar \Sigma}^{R}_{+}(1,\bar\varepsilon)/
d\ln \bar\varepsilon$ is shown in the insets.
%Figs.~\ref{fig:2}~(d) and~(e). 
At low energies,
$\zeta(\bar\varepsilon)\to 2$ consistently with FL behavior. Conversely, above the crossover energy it approaches the NFL value
$\zeta(\bar\varepsilon)\simeq 2/3$.

This scaling closely parallels that of fermions coupled to overdamped critical bosons~\cite{Polchinski1994,altshuleroffeMills1994,Lee2009GaugeNFL,MetlitskiSachdev2010,Mross2010,maslov2017optical,GindikinMaslovChubukov2025}. There, Landau damping yields an effective bosonic dynamical exponent $z=3$ and the standard one-loop result $\mathrm{Im}\Sigma(\varepsilon)\sim \varepsilon^{2/3}$, signaling the loss of quasiparticle coherence. In our case, however, the bosonic mode is not a conventional order-parameter fluctuation near a quantum critical point, but a gauge-like acoustic phonon.
A related $\varepsilon^{2/3}$ self-energy was reported in~\cite{rao2023non} for electrons coupled to a critical transverse electromagnetic cavity mode at the superradiant transition. The distinction here is that the dressed acoustic phonon propagator, Eq.~\eqref{eq:D_TT_final}, contains both dispersive and dissipative terms. The overdamped regime appears only above the crossover scale $\bar\varepsilon_*$, where dissipation exceeds the bare dispersion. Accordingly, Fig.~\ref{fig:2} shows a crossover from low-energy FL behavior to $\mathrm{Im}\Sigma(\bar\varepsilon)\sim \bar\varepsilon^{2/3}$ for $\bar\varepsilon>\bar\varepsilon_*$. We note that analyses beyond the one-loop approximation find that NFL behavior persists \cite{Rechetal}. However, functional-renormalization-group studies have reported critical exponents differing from conventional expectations in two-dimensional quantum-critical metals~\cite{HooleyPRL2015,HooleyPRB2018} and for electrons coupled to a $U(1)$ gauge field~\cite{HooleyPRR2025}. %{\color{blue}(Marco: ok, yes, but I would make it clear that the NFL behavior is robust and not washed out by theory beyond one-loop)}

Figure~\ref{fig:2}(c) shows the self-energy for a paramagnetic $\bar\chi_{\rm orb}>0$, which softens the dressed phonon dispersion ${\bar \omega}_{{\rm ph},q}$. 
While orbital paramagnetism can lead to a superradiant phase transition~\cite{guerci2020superradiant,rao2023non,andolina2020photon} %{\color{blue}(you totally miss to cite the paper by Andolina and myself in PRB 2020 together with Daniele's PRL: https://doi.org/10.1103/PhysRevB.102.125137)}, 
in the overdamped regime a qualitatively different scaling of the self-energy emerges regardless of the existence of an instability.

In contrast to the diamagnetic case, the system behaves as a MFL in the infrared regime. The logarithmic derivative approaches $\zeta(\bar\varepsilon)\simeq 1$ as $\bar\varepsilon\to 0$, corresponding to $\mathrm{Im}\Sigma^{R}_{+}(\bar\varepsilon)\propto\bar\varepsilon$, while at higher energies the self-energy retains the NFL scaling $\propto \bar\varepsilon^{2/3}$ associated with the Landau-damped mode. Notably~\footnotemark[\value{SOM}], the crossover scale $\bar\varepsilon_*$ continues to govern the transition
between the two regimes. 
The MFL infrared regime originates from the presence of a ``shell'' of softened phonon modes~\footnotemark[\value{SOM}] at the characteristic finite momentum
$q_{\rm shell}\sim {\bar c}_{\rm ph}/\sqrt{\bar\chi_{\rm orb}}$. This is reminiscent of Brazovskii-type soft-mode physics~\cite{brazovskii1975nonuniform},
in which the bosonic spectrum softens at finite momentum, thereby enhancing the phase space of fluctuations. 
%In this
%respect, the paramagnetic case is also analogous to a finite-$q$ instability,
%such as a charge-density-wave problem, but with the important caveat that
%the softened modes lie on a shell of finite momenta rather than at a single
%ordering wave vector.

%The physical content of this regime is distinct from the conventional
%phenomenological marginal-Fermi-liquid picture. 
%In the original proposal of
%Varma \textit{et al.}~\cite{varma1989marginal}, the linear-in-energy
%scattering rate was introduced phenomenologically in terms of a broad
%critical bosonic spectrum. Here, by contrast, the marginal-Fermi-liquid
%scaling emerges microscopically when $\bar\chi_{\rm orb}>0$. 
Figure~\ref{fig:2}(c), together with the analytical asymptotic behavior~\footnotemark[\value{SOM}]
$\mathrm{Im}\Sigma^{R}_{+}(\bar\varepsilon)\propto\bar\varepsilon$, shows that the present model 
%does not merely
%reproduce a marginal-Fermi-liquid exponent. Rather, it 
provides a microscopic route
to MFL behavior~\cite{varma1989marginal}.
%together with a crossover into a non-Fermi-liquid regime, with the crossover occurring at the same scale,
%$\bar\varepsilon_{\ast}$~.
We emphasize that the MFL and NFL contributions compete with conventional FL behavior arising from electron-electron interactions (for which $\mathrm{Im}\Sigma(\varepsilon)\sim \varepsilon^2$~\cite{ChubukovMaslov}). However, both the MFL and NFL self-energies are parametrically larger, with the former dominating in the infrared regime and the latter for $\varepsilon \gtrsim \varepsilon_*$.

We now turn to estimates for realistic Dirac systems. Specifically, we consider monolayer graphene and an analogous Dirac system with a reduced Fermi velocity. Since the latter may have implications for MATBG, we refer to it as such in what follows. However, we emphasize that quantitative estimates for MATBG would require incorporating microscopic details that lie beyond the scope of the modelling performed in this paper.
In both cases, phonon and electron-phonon parameters are taken to be those of graphene. In graphene the phonon velocities~\cite{kaasbjerg2012unraveling,Sohier2014PRB_GaugePhononResistivity} are, $c_{{\rm ph},{\rm L}}=2.2\times10^4~{\rm m\,s}^{-1}$ and $c_{{\rm ph},{\rm T}}=1.4\times10^4~{\rm m\,s}^{-1}$. The mass density and the electron-phonon coupling constant are $\rho_m = 7.6\times 10^{-8}~{\rm g\,cm}^{-2}$ \cite{castro2009electronic} and $g_2 \approx 1.5~{\rm eV}$ \cite{Suzuura,Mariani,chen}, respectively. 
The mass density of MATBG is twice that of monolayer graphene~\cite{castro2009electronic}, but this factor is compensated by the extra minivalley degeneracy in the definition of ${\bar \chi}_{\rm orb}$ and ${\bar \alpha}$.
% To connect the effective theory to experiment, we estimate the parameters entering the dressed transverse phonon propagator in Eq.~(\ref{eq:D_TT_final}) from the transverse current response. 
% For the static term, we use the relation~\cite{gomez2011measurable}
% $\bar\chi=\Re(\chi_{\rm orb})/\mu_0 e^2$, where $\chi_{\rm orb}$ is the orbital
% magnetic susceptibility. 

\begin{figure}
\includegraphics[width=0.48\textwidth]{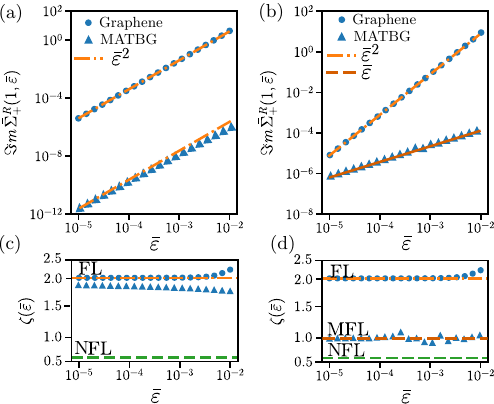}
\caption{
\label{fig:3}
The imaginary part of the self-energy for monolayer graphene and a minimal model of MATBG. Material parameters are given in the main text. 
(a) and (c) show the diamagnetic case. The dashed line is a guide to the eye showing FL scaling $\propto \varepsilon^2$. 
%For monolayer graphene (blue), the self-energy follows the Fermi liquid behavior over the entire window. In contrast to 
The self-energy of MATBG (diamond) shows a departure from FL behavior at energies much smaller than $\bar{\varepsilon}_{\ast}\sim 10$.
%enhanced electronic response leads to an observable departure from Fermi-liquid behavior, with panel (c) showing the logarithmic derivative in both cases. 
%Panels 
(b) and (d) show the paramagnetic case. 
%The solid line is a guide to the eye showing MFL scaling $\propto \varepsilon$. 
For monolayer graphene, the self-energy still follows the FL behavior, while 
%owing to inaccessible $q_{\rm shell}$. In contrast, 
in MATBG the enhanced electronic response leads to MFL behavior
over the entire energy window.
%owing to an accessible $q_{\rm shell}$. 
%Panel (d) shows the logarithmic derivative for both cases, with MATBG exhibiting marginal Fermi-liquid-like behavior.
}
\end{figure}

We begin by considering the diamagnetic regime ($\chi_{\rm orb}<0$) in Figs.~\ref{fig:3}(a) and \ref{fig:3}(c).
%Since the magnetic orbital response of undoped Dirac electrons is intrinsically diamagnetic~\cite{McClure1956}, 
This behavior applies to both undoped or weakly doped monolayer graphene~\cite{McClure1956} and MATBG~\cite{Guercidiamagnetic}.
% in regimes where a Dirac band dispersion dominates the orbital response.
%
%To estimate the magnitude of the orbital response, 
For graphene, we use~\cite{gomez2011measurable} $ \bar\chi_{\rm orb}= -1.72 \times 10^{-8}$
and~\cite{principi2009linear} 
%, obtained from lattice estimates of the magnetic susceptibility. 
%We also use the imaginary part of the leading-order transverse current-current response of doped graphene
%~\cite{principi2009linear} 
%to estimate 
$\bar \alpha = 7\times10^{-8}$ at $\varepsilon_{\rm F}=0.1\,{\rm eV}, v_{\rm F} = 10^{6}~{\rm m\,s^{-1}}$. The details of the estimates of $\bar\chi_{\rm orb}$ and $\bar\alpha$
%, based on the transverse current-current response, 
are provided in~\footnotemark[\value{SOM}].
% Appendix~\ref{app:parameters}.
%
Values for MATBG are obtained by replacing the Fermi velocity and energy with~\cite{morell2010flat} $v_{\rm F} = 10^4\ {\rm m\,s^{-1}}$ and $\varepsilon_{\rm F}= 1\ {\rm meV}$. 
%Within the same procedure (see Appendix~\ref{app:parameters}), 
This leads~\footnotemark[\value{SOM}] to the dimensionless parameters $\bar\chi_{\rm orb}=-1.72$ and $\bar\alpha = 7\times10^{-2}$, indicating a substantial enhancement compared to graphene.
%, after rescaling,  primarily driven by the strong reduction of $v_{\rm F}$ in MATBG.
%
In both systems the low-energy behavior remains FL-like, with $\mathrm{Im}\,\bar{\Sigma}^{R}_{+}\propto \bar{\varepsilon}^{2}$. For graphene, this behavior persists throughout the entire energy window shown, consistent with the estimate $\bar{\varepsilon}_{\ast}\sim 10^{5}$, which lies far above the plotted range. Although in MATBG the threshold energy $\bar{\varepsilon}_{\ast}\sim 10$ is still too large to observe a marked NFL behavior in the chosen energy window, 
%the enhanced dimensionless parameters $\bar{\chi}_{\rm orb}$ and $\bar{\alpha}$ already lead to 
%a visible
a departure from FL behavior is already visible at relatively small energies.
%at higher energies, as seen in Figures~\ref{fig:3}(a) and \ref{fig:3}(c).

Figures~\ref{fig:3}(b) and \ref{fig:3}(d) show the paramagnetic case, $\chi_{\rm orb}>0$. 
Orbital paramagnetism can arise from various mechanisms, including interactions~\cite{principiparamagnetisim}, multiband and flat-band physics~\cite{flatbandparamagentisim}, and proximity to van Hove singularities~\cite{vhsparamagentisim}. Here, we use an interaction-induced~\cite{principiparamagnetisim} $\bar{\chi}_{\rm orb} = 2.6\times10^{-8}$ for graphene and a substaintially enhanced $\bar{\chi}_{\rm orb}= 2.6$ for MATBG~\footnotemark[\value{SOM}]. %{\color{blue}(Marco: wow! Eight orders of magnitude! Shall we emphasize that? When one reads it fast, it looks just like a typo, you know what I mean????)}%
We find that, for graphene, the self-energy remains FL-like because the soft-shell momentum lies well outside the integration domain. In contrast, for MATBG, it enters the momentum integration range, producing a dominant contribution from softened finite-$q$ modes. As a result, the system exhibits MFL scaling, $\mathrm{Im} \bar{\Sigma}^{R}_{+}\propto \bar{\varepsilon}$ over the plotted energy window. The logarithmic derivative confirms these scalings. %{\color{blue}(Marco: I remember asking this to Alessandro long time ago. Is there any literature showing that this leads to Planckian resistivity? Second, more important question: usually when phonon-mechanisms are involved, one gets linear-in-$T$ resistity not down to zero temperature but only down to the large Bloch-Gruneisen temperature $T_{\rm BG}$ scale, which here did not appear. What is the role of $T_{\rm BG}$ here?)}

\textit{Conclusions---}In this work, we showed that strain-induced gauge phonons, which couple to electronic currents rather than densities, can drive NFL behavior in Dirac materials without requiring proximity to a quantum critical point. 
% We highlight MATBG as a promising platform, while noting that a full treatment of microscopic details would be required to draw definitive conclusions for a realistic system.
% The key control parameters are the orbital susceptibility $\chi_{\rm orb}$ and the phonon damping strength $\bar{\alpha}$ due to coupling to transverse currents. When phonons are strongly overdamped, $\bar{\alpha} \gg 1$, they produce robust departures from FL theory.
% 
% The sign of $\chi_{\rm orb}$ determines the low-energy behavior. For diamagnetic systems, $\chi_{\rm orb}<0$, a FL regime survives only in a very narrow low-energy window before crossing over to NFL scaling. For paramagnetic systems, $\chi_{\rm orb}>0$, the FL regime disappears entirely: softened transverse phonons instead generate MFL behavior at low energies, followed by NFL scaling at higher energies. We stress that the MFL regime we discuss here arises at finite doping, where electron-electron interactions alone would yield an ordinary FL. Therefore, it is distinct from the one occurring in undoped Dirac systems. {\color{blue}(Marco: a summary in the conclusions section is neither required not typical in PRL these days. Also because this very same summary has appeared a zillion times, starting from the abstract....)}
% 
% More broadly, 
The same mechanism may be relevant in systems with different lattice geometries, since the existence of gauge phonons that couple to electronic currents does not rely on hexagonal symmetry alone. 
% Similar gauge-like phonon couplings can arise in other non-square lattices. 
% {\color{blue}(Marco: again, also this statement just above sounds like another repetition.)} 
% It
A similar gauge-like phonon coupling
may also occur in distorted square-lattice systems where pseudo-gauge fields emerge once symmetry is reduced~\cite{sun2023square}. In this sense, the present mechanism may offer a potentially generic microscopic route towards NFL phenomenology 
% {\color{blue}(Marco: careful, we never showed strange metallicity since we did not do any calculation of resistivity. We just showed NFL windows.)}
in a very broad class of materials. Assessing its possible relevance to systems such as the cuprates, where linear-in-temperature transport remains a central open problem~\cite{Bruin2013,Legros2019} and is sometimes related to a NFL behavior of the self-energy \cite{phillips2022stranger}, would require a more detailed investigation that is beyond the scope of the present work.

\textit{Acknowledgments.—}AP, MSB and AK acknowledge the support of the Leverhulme Trust under Grant Agreement RPG-2023-253. 
RG acknowledges the support of the Departmental Studentship funded by the Faculty of
Science and Engineering, University of Manchester.
MP was partly supported by the European Union under grant agreement No. 101131579 - Exqiral. 
Views and opinions expressed in this material are those of the author(s) only and do not necessarily reflect those of the European Union or the European Commission. 
Neither the European Union nor the granting authority can be held responsible for them.

\section*{End Matter}

\subsection*{Comparison with existing mechanisms for non-Fermi-liquid behavior}

The mechanism proposed here is close in spirit to theories of fermions coupled to transverse gauge fields~\cite{Polchinski1994,AltshulerIoffeMillis1994,Lee2009GaugeNFL,LeeNagaosaWen2006}. In both cases, the boson couples to electronic currents, the transverse response contains a regular $q^2$ term and a Landau-damping term, and strong damping can produce the characteristic two-dimensional scaling $\Gamma(\varepsilon)\propto\varepsilon^{2/3}$. The crucial difference is that the transverse mode in our problem is a lattice vibration with gauge-phonon character~\cite{Sohier2014PRB_GaugePhononResistivity}, rather than an emergent gauge field or a $U(1)$ gauge boson. As such, it couples to valley-antisymmetric (rather than valley-symmetric) currents\footnote{Note however that, since the self-energy depends on the coupling squared, and we neglect inter-valley scattering, there is no cancellation of contributions coming from different valleys.}.

It also differs from order-parameter quantum criticality. In Hertz--Millis-type theories~\cite{Hertz1976,Millis1993}, singular scattering is tied to soft critical fluctuations near a continuous phase transition, with exponents and crossovers depending on the universality class~\cite{Lohneysen2007,ChubukovReview,MetlitskiSachdev2010,Mross2010}. Functional-renormalisation-group studies have shown that two-dimensional quantum-critical metals can display anomalous damping and non-Fermi-liquid fixed points~\cite{HooleyPRL2015,HooleyPRB2018}, while related $U(1)$ gauge-field problems may exhibit different scaling structures~\cite{HooleyPRR2025}. By contrast, the gauge-phonon route is controlled by the orbital response $\chi_{\rm orb}$ and the damping strength $\bar\alpha$, rather than by tuning to an order-parameter critical point.

The present work is also distinct from marginal-Fermi-liquid phenomenology~\cite{varma1989marginal} and from mechanisms in which randomness plays an essential role, such as random-Yukawa models producing robust strange-metal transport~\cite{Patel2023Universal}. Here the marginal regime appears dynamically in the paramagnetic case, $\chi_{\rm orb}>0$, because softening of the gauge-phonon dispersion removes the conventional FL infrared behavior. Finally, while cavity-QED electron gases provide another current-coupled setting~\cite{rokaj2020free}, and cavity fluctuations near superradiance can generate NFL behavior~\cite{rao2023non}, the present mechanism is intrinsic to crystalline systems with suitable lattice geometry~\cite{sun2023square}. Dirac materials, where strain acts as a valley-dependent gauge field~\cite{Vozmediano2010,Guinea2010NatPhys}, offer a natural setting. The resulting marginal behavior extends earlier discussions of interaction effects in charge-neutral Dirac systems~\cite{das2007many,barlas2009non} to doped systems that would otherwise be conventional Fermi liquids. Our work therefore complements broader non-quasiparticle perspectives on strange metals, including SYK-inspired approaches~\cite{chowdhury2022sachdev}.

\subsection*{Singularities and asymptotics of $\Xi(\bar{q},\bar{\omega},\bar{\varepsilon})$ }

The expression for $\Xi(\bar{q},\bar{\omega},\bar{\varepsilon})$ alluded to in the main text is given as 

\begin{equation}
    \Xi(\bar{q},\bar{\omega},\bar{\varepsilon}) = \frac{\bar{q} |\bar{\varepsilon} + \bar{\omega}+ 1|}{\sqrt{(2 \bar{q})^2 - \big[(\bar{\varepsilon} + \bar{\omega} + 1)^2 -(1+\bar{q}^2)\big]^2}}.
\end{equation}

The function $\Xi(\bar{q},\bar{\omega},\bar{\varepsilon})$ is smooth throughout the interior of the integration domain and, being dimensionless, does not affect the energy scaling of the self energy. Although it exhibits square-root like divergences at the integration boundaries, these are integrable endpoint singularities rather than true divergences \cite{Duffy1982}. 

In the Fermi liquid regime, where the dominant momenta scale as $\bar q \sim \bar\varepsilon,$ one sets $\bar\omega=\tilde\omega\,\bar\varepsilon$ and $\bar q=\tilde q\,\bar\varepsilon$. Then
$
\bar\varepsilon+\bar\omega+1 = 1+\mathcal O(\bar\varepsilon)$ and $
1+\bar q^2 = 1+\mathcal O(\bar\varepsilon^2),
$
so that
$
(2\bar q)^2-\left[(\bar\varepsilon+\bar\omega+1)^2-(1+\bar q^2)\right]^2
=
4\bar\varepsilon^2\left[\tilde q^2-(1+\tilde\omega)^2\right]+\mathcal O(\bar\varepsilon^3),
$
and hence
\begin{equation}
\Xi(\bar q,\bar\omega,\bar\varepsilon)
\to
\frac{\tilde q}{2\sqrt{\tilde q^2-(1+\tilde\omega)^2}}.
\end{equation}

Thus, in the Fermi liquid regime $\Xi(\bar q,\bar\omega,\bar\varepsilon)$ reduces to a finite dimensionless function of the rescaled variables and does not generate additional powers of $\bar\varepsilon.$

In the non Fermi liquid regime, where the dominant momenta scale as $\bar q\sim \bar\varepsilon^{1/3},$ one instead sets $\bar\omega=\tilde\omega\,\bar\varepsilon$ and $\bar q=\tilde q\,\bar\varepsilon^{1/3}$. In this case,
$
\bar\varepsilon+\bar\omega+1 = 1+\mathcal O(\bar\varepsilon)$ and
$1+\bar q^2 = 1+\mathcal O(\bar\varepsilon^{2/3})
$.
Thus, to leading order we have $\bar{q}|1+\bar{\omega}+\bar{\varepsilon}| \sim1 +O(\bar{\varepsilon}^{1/3})$. The denominator scales similarly as, 
$
(2\bar q)^2-\left[(\bar\varepsilon+\bar\omega+1)^2-(1+\bar q^2)\right]^2
=
4\bar\varepsilon^{2/3}\tilde q^2+\mathcal O(\bar\varepsilon^{4/3}),
$
which implies
$
\Xi(\bar q,\bar\omega,\bar\varepsilon)\to 1/2,
$
up to subleading corrections as $\bar\varepsilon\to 0.$ Therefore, in both scaling limits, $\Xi(\bar q,\bar\omega,\bar\varepsilon)$ contributes only an overall order one factor. At the same time, the bosonic kernel entirely controls the energy dependence of the self-energy.

\clearpage

\clearpage
\onecolumngrid

\setcounter{equation}{0}

\renewcommand{\theequation}{S\arabic{equation}}
\begin{center}
{\large\bfseries Supplemental Online Material for ``Gauge phonon driven non-Fermi liquids in Dirac materials''\par}
\end{center}

\vspace{2em}

\twocolumngrid

\section{Derivation of the strain-induced electron--phonon coupling in graphene}
\label{app:strain}

In this appendix, we summarise the standard derivation of the low-energy scalar and vector electron--phonon couplings used in Eq.~(\ref{eq:e_ph_ham}), following Refs.~\cite{Vozmediano2010PhysRep,Guinea2010NatPhys,de2012space,castro2009electronic}. We start from the modulation of nearest-neighbour hoppings by a smooth in-plane displacement field ${\bm u}({\bm r})$. Our goal is to recover the electron-phonon coupling Hamiltonian given in the main text, 
\begin{equation}
{\cal H}_{\rm e\text{-}ph}
=
\sum_{{\bm k},{\bm q},\alpha,\beta}
c^\dagger_{{\bm k}-{\bm q},\alpha}
\Big({\bm A}_{\bm q}\!\cdot\!{\bm j}_{\bm q}
+V_{\bm q}\openone\Big)_{\alpha\beta}
c_{{\bm k},\beta},
\end{equation}
where $V_{\bm q}$ is a scalar potential describing the
deformation-potential coupling, while 
${\bm A}_{{\bm q}} = (A_{{\bm q},{\rm L}},A_{{\bm q},{\rm T}})$ is the strain-induced gauge potential. For later convenience, we decompose it in components parallel and perpendicular to ${\bm q}$, {\it i.e.} $A_{{\bm q},{\rm L}} = {\hat {\bm q}}\cdot {\bm A}_{{\bm q}}$ and $A_{{\bm q},{\rm T}} = ({\hat {\bm z}}\times {\hat {\bm q}})\cdot {\bm A}_{{\bm q}}$, respectively.
As such, the current operator ${\bm j}_{\bm q} = (j_{{\bm q},{\rm L}}, j_{{\bm q},{\rm T}})$ is also decomposed into its longitudinal and transverse current
components
$j_{{\bm q},{\rm L}}=v_{\rm F}\,\hat{\bm q}\!\cdot\!{\bm \sigma}$ and
$j_{{\bm q},{\rm T}}=v_{\rm F}\,(\hat{\bm z}\times\hat{\bm q})\!\cdot\!{\bm \sigma}$, respectively.
Here, ${\bm \sigma}=(\sigma^x, \sigma^y)$ is a two-component vector of Pauli matrices acting on the sublattice degree of freedom. 

The scalar potential can be written in terms of the phonon creation and annihilation operators of mode $\nu={\rm L},{\rm T}$, {\it i.e.} $a^\dagger_{{\bm q},\nu}$ and $a_{{\bm q},\nu}$, respectively, as
\begin{equation}
V_{\bm q}
=
g_1\,q
\sqrt{\frac{1}{2\rho_m\omega_{{\bm q},{\rm L}}}}\,
\Big(a_{{\bm q},{\rm L}}+a^\dagger_{-{\bm q},{\rm L}}\Big),
\label{eq:Vq_app_summary}
\end{equation}
while the gauge potential takes the form
\begin{equation}
A_{{\bm q},\nu}
=
\sum_{\nu'={\rm L,T}}
g_{\nu'}(q)\,
 M_{\nu\nu'}(\hat{\bm q})\,
\Big(a_{{\bm q},\nu'}+a^\dagger_{-{\bm q},\nu'}\Big),
\label{eq:Aq_app_summary}
\end{equation}
with
\begin{equation}
g_{\nu}(q)
=
\frac{g_2}{v_{\rm F}}
\sqrt{\frac{q^2}{2\rho_m\omega_{{\bm q},\nu}}},
\qquad
 M(\hat{\bm q})
=
\begin{pmatrix}
\cos\varphi_{\bm q} & -\sin\varphi_{\bm q}\\
\sin\varphi_{\bm q} & \cos\varphi_{\bm q}
\end{pmatrix}.
\label{eq:gq_Mtilde_app_summary}
\end{equation}
We now summarise the derivation of these expressions from the strain dependence of the
nearest-neighbour hoppings.

Let ${\bm \delta}_n$ ($n=1,2,3$) denote the undeformed nearest-neighbour
vectors, with $|{\bm \delta}_n|=a$ and
$\hat{\bm \delta}_n={\bm \delta}_n/a$. To linear order in gradients, the
deformed bond vectors are
\begin{equation}
{\bm \delta}'_n({\bm r})
=
{\bm \delta}_n+{\bm u}({\bm r}+{\bm \delta}_n)-{\bm u}({\bm r})
\simeq {\bm \delta}_n+({\bm \delta}_n\cdot {\bm \nabla}){\bm u}.
%(\mathbb{I}+\nabla{\bm u})\,{\bm \delta}_n .
\end{equation}
The corresponding fractional bond-length change is
\begin{equation}
\frac{|{\bm \delta}'_n|-a}{a}
\simeq
\hat{\bm \delta}_n^{\,T}\,u({\bm r})\,\hat{\bm \delta}_n,
\qquad
u_{ij}({\bm r})=\frac{1}{2}\big(\partial_i u_j+\partial_j u_i\big),
\end{equation}
where $u_{ij}$ is the strain tensor.
Modelling the strain-induced hopping modulation as
\cite{Vozmediano2010,de2012space}
\begin{equation}
t_n({\bm r})
=
t\exp\!\left[
-\beta\left(\frac{|{\bm \delta}'_n({\bm r})|}{a}-1\right)
\right]
\simeq
t+\delta t_n({\bm r}),
\end{equation}
gives, to linear order in $u_{ij}({\bm r})$,
\begin{equation}
\delta t_n({\bm r})
\simeq
-\beta t\,\hat{\bm \delta}_n^{\,T}\,u({\bm r})\,\hat{\bm \delta}_n.
\end{equation}
For graphene one has $\beta\simeq 3$ \cite{Vozmediano2010PhysRep}.

Near the valley ${\bm K}_\tau$ with $\tau=\pm1$, we define
\begin{equation}
\delta f_\tau({\bm r})
\equiv
\sum_{n=1}^3
\delta t_n({\bm r})\,e^{i{\bm K}_\tau\cdot{\bm \delta}_n}.
\end{equation}
For the standard nearest-neighbour vectors one finds
\cite{Vozmediano2010PhysRep,de2012space}
\begin{equation}
\delta f_{\tau}({\bm r})
=
\tau\frac{3\beta t}{4}
\Big[(u_{xx}-u_{yy})-i\,2u_{xy}\Big].
\label{eq:df_xi_app}
\end{equation}
This correction enters the Dirac Hamiltonian as a valley-odd pseudo-gauge
field. Writing $\Psi_\tau({\bm k})$ for the two-component pseudospin spinor in
valley $\tau$, the strain contribution takes the form
\begin{equation}
{\cal H}_{\rm strain}
=
v_{\rm F}
\sum_{\tau=\pm1}
\sum_{{\bm k},{\bm q}}
\Psi^\dagger_{\tau}({\bm k}-{\bm q})\,
\big[\tau\,{\bm A}({\bm q})\cdot{\bm \sigma}\big]\,
\Psi_{\tau}({\bm k}),
\label{eq:Hstrain_kspace}
\end{equation}
so that the corresponding current operator is
${\bm j}=\delta H/\delta{\bm A}=v_{\rm F}{\bm \sigma}$. The real-space
pseudo-gauge field is therefore
\begin{equation}
{\bm A}({\bm r})
=
g_2
\begin{pmatrix}
u_{xx}({\bm r})-u_{yy}({\bm r})\\
-2u_{xy}({\bm r})
\end{pmatrix},
\qquad
g_2=\frac{3\beta t}{4}.
\label{eq:A_of_r_app}
\end{equation}

We now express ${\bm A}({\bm q})$ in terms of phonon modes by quantizing the
displacement field as
\begin{align}
&{\bm u}({\bm r})
=
\frac{1}{\sqrt{N}}
\sum_{{\bm q},\nu={\rm L,T}}
\sqrt{\frac{1}{2\rho_m\omega_{{\bm q},\nu}}}\,
{\bm e}_{\nu}(\hat{\bm q}) \nonumber \\
&\times \Big(
a_{{\bm q},\nu}e^{i{\bm q}\cdot{\bm r}}
+
a^\dagger_{-{\bm q},\nu}e^{-i{\bm q}\cdot{\bm r}}
\Big),
\label{eq:u_quant_app}
\end{align}
where ${\bm e}_{\rm L}=\hat{\bm q}$ and
${\bm e}_{\rm T}=\hat{\bm z}\times\hat{\bm q}$. Since
$u_{ij}({\bm q})=\frac{i}{2}(q_i u_j+q_j u_i)$, the combinations entering
Eq.~(\ref{eq:A_of_r_app}) are
\begin{eqnarray}
u_{xx}-u_{yy} &=& i(q_xu_x-q_yu_y), \\
-2u_{xy} &=& -i(q_xu_y+q_yu_x).
\label{eq:uxx_uyy_uxy_app}
\end{eqnarray}
Writing the displacement in the polarization basis
${\bm u}=u_{\rm L}\hat{\bm q}+u_{\rm T}(\hat{\bm z}\times\hat{\bm q})$,
one has
$u_x=\cos\varphi_{\bm q}\,u_{\rm L}-\sin\varphi_{\bm q}\,u_{\rm T}$ and
$u_y=\sin\varphi_{\bm q}\,u_{\rm L}+\cos\varphi_{\bm q}\,u_{\rm T}$.
Projecting the gauge field onto longitudinal and transverse components,
$A_{{\bm q},{\rm L}}=\hat{\bm q}\cdot{\bm A}({\bm q})$ and
$A_{{\bm q},{\rm T}}=(\hat{\bm z}\times\hat{\bm q})\cdot{\bm A}({\bm q})$,
one obtains
\begin{eqnarray}
A_{{\bm q},{\rm L}}
&=&
g_2 q
\Big(
\cos\varphi_{\bm q}\,u_{\rm L}
-
\sin\varphi_{\bm q}\,u_{\rm T}
\Big), \\
A_{{\bm q},{\rm T}}
&=&
g_2 q
\Big(
\sin\varphi_{\bm q}\,u_{\rm L}
+
\cos\varphi_{\bm q}\,u_{\rm T}
\Big).
\label{eq:A_LT_from_u_app}
\end{eqnarray}
Here the overall factor of $i$ originating from gradients has been absorbed
into the phase convention for the phonon operators, so that
$A_{{\bm q},\nu}$ is chosen real. Using
$u_{\nu}({\bm q})=\sqrt{1/(2\rho_m\omega_{{\bm q},\nu})}\,
(a_{{\bm q},\nu}+a^\dagger_{-{\bm q},\nu})$,
Eq.~(\ref{eq:A_LT_from_u_app}) reduces to
Eq.~(\ref{eq:Aq_app_summary}) with
$g_\nu(q)$ and $M(\hat{\bm q})$ given in
Eq.~(\ref{eq:gq_Mtilde_app_summary}).

Finally, the scalar deformation-potential coupling follows from
$V({\bm r})=g_1\big(u_{xx}+u_{yy}\big)$ \cite{castro2009electronic}. Using Eq.~(\ref{eq:u_quant_app}),
one finds
\begin{align}
&u_{xx}({\bm q},\nu)+u_{yy}({\bm q},\nu)
=
iq\sqrt{\frac{1}{2\rho_m\omega_{{\bm q},\nu}}}\ \nonumber \\
&\times \big(
a_{{\bm q},\nu}+a^\dagger_{-{\bm q},\nu}
\big)
({\bm e}_{\nu}\cdot\hat{\bm q}),
\end{align}
so that only longitudinal phonons contribute $V({\bm r})$. This yields
Eq.~(\ref{eq:Vq_app_summary}),
\begin{equation}
V_{\bm q}
=
g_1\,q
\sqrt{\frac{1}{2\rho_m\omega_{{\bm q},{\rm L}}}}\,
\Big(
a_{{\bm q},{\rm L}}+a^\dagger_{-{\bm q},{\rm L}}
\Big).
\label{eq:Vq_app}
\end{equation}

For completeness, the explicit gauge-field coupling entering the main text is
therefore
\begin{equation}
A_{{\bm q},\nu}
=
\sum_{\nu'={\rm L,T}}
g_{\nu'}(q)\, M_{\nu\nu'}(\hat{\bm q})\,
\Big(a_{{\bm q},\nu'}+a^\dagger_{-{\bm q},\nu'}\Big),
\end{equation}
with
\begin{equation}
\label{eq:gaugephonons}
g_{\nu}(q)
=
\frac{g_2}{v_{\rm F}}
\sqrt{\frac{q^2}{2\rho_m\omega_{{\bm q},\nu}}},
~
M(\hat{\bm q})
=
\begin{pmatrix}
\cos\varphi_{\bm q} & -\sin\varphi_{\bm q}\\
\sin\varphi_{\bm q} & \cos\varphi_{\bm q}
\end{pmatrix}.
\end{equation}

\section{Self-energy derivation}
\label{app:self_energy}

In this appendix, we provide the intermediate steps leading from Eq.~\eqref{eq:SE_def} to the transverse-only form
used in the low-energy analysis.

We begin by recalling that the dressed phonon propagator is defined as the two-point
correlator of the vector-potential components
\begin{align}
&D^{\nu\nu'}({\bm q},i\omega_m)
=
-\int_0^\beta d\tau\,e^{i\omega_m\tau}\, \nonumber \\
&\times \big\langle {\cal T}_\tau  
A_{{\bm q},\nu}(\tau)A_{-{\bm q},\nu'}(0)\big\rangle,
\label{eq:D_def}
\end{align}
where $i\omega_m$ is a bosonic Matsubara frequency.

The bare electron Green's function is
\begin{equation}
G^{(0)}_{\lambda}({\bm k},i\varepsilon_n)
=
\frac{1}{i\varepsilon_n-\xi_{{\bm k},\lambda}},
\end{equation}
where $i\varepsilon_n$ is a fermionic Matsubara frequency.

For later use, the Cartesian current matrix elements between incoming and
outgoing Dirac states are
\begin{align}
j^{x}_{{\bm k},\lambda;{\bm k}',\lambda'}
&=
v_{\rm F}\frac{\lambda e^{-i\varphi_{\bm k}}
+\lambda' e^{i\varphi_{{\bm k}'}}}{2},
\\
j^{y}_{{\bm k},\lambda;{\bm k}',\lambda'}
&=
v_{\rm F}\frac{i\lambda e^{-i\varphi_{\bm k}}
-i\lambda' e^{i\varphi_{{\bm k}'}}}{2}.
\label{eq:app_jxy}
\end{align}

Since we need the retarded self-energy, we analytically continue
$i\varepsilon_n\to\varepsilon+i0^+$. Starting from Eq.~\eqref{eq:SE_def},
we rewrite the Matsubara sum as a contour integral,
\begin{align}
&\Sigma_\lambda({\bm k},i\varepsilon_n)
=
-\oint\frac{dz}{2\pi i}\,n_{\rm B}(z)
\sum_{\nu,\nu',\lambda'}
\int\frac{d^2{\bm q}}{(2\pi)^2}
\nonumber\\
&\times
j^{\nu}_{{\bm k},\lambda;{\bm k}+{\bm q},\lambda'}
j^{\nu'}_{{\bm k}+{\bm q},\lambda';{\bm k},\lambda}
D^{\nu,\nu'}({\bm q},z)
G^{(0)}_{\lambda'}({\bm k}+{\bm q},i\varepsilon_n+z).
\label{eq:app_sigma_analit_1}
\end{align}
with the contour encircling the poles of $n_{\mathrm{B}}(z)=(e^{\beta z}-1)^{-1}$ anticlockwise. Here, $D^{\nu,\nu'}({\bm q},z)$ and $G^{(0)}_{\lambda}({\bm k},z)$ are understood as the phonon and electron
propagators analytically continued away from the imaginary-frequency axis. The
integrand has branch cuts at $\mathrm{Im}(z)=0$ and $\mathrm{Im}(z)=-i\varepsilon_n$.
The contour in Eq. \eqref{eq:app_sigma_analit_1} can be deformed to circle these clockwise. The resulting integral after analytically  continuing
$i\varepsilon_n\to\varepsilon+i0^+$ yields
\begin{align}
&\Sigma_\lambda^{(R)}({\bm k},\varepsilon)
=
-\int_{-\infty}^{\infty}\frac{d\omega}{\pi}
\sum_{\nu,\nu',\lambda'}
\int\frac{d^2{\bm q}}{(2\pi)^2}
j^{\nu}_{{\bm k},\lambda;{\bm k}+{\bm q},\lambda'}
j^{\nu'}_{{\bm k}+{\bm q},\lambda';{\bm k},\lambda}
\nonumber\\
&\times
\Big[
n_{\rm B}(\omega)\,
\mathrm{Im} D^{\nu,\nu'}({\bm q},\omega+i\eta)\,
G^{(0)}_{\lambda'}({\bm k}+{\bm q},\varepsilon+\omega+i\eta)
\nonumber\\
&-
n_{\rm F}(\omega+\varepsilon)\,
D^{\nu,\nu'}({\bm q},\omega-i\eta)\,
\mathrm{Im} G^{(0)}_{\lambda'}({\bm k}+{\bm q},\varepsilon+\omega+i\eta)
\Big],
\label{eq:app_sigma_analit_2}
\end{align}
with $\eta\to0^+$. Taking the imaginary part gives
\begin{align}
&\mathrm{Im} \Sigma^R_\lambda({\bm k},\varepsilon)
=
-\int_{-\infty}^{\infty}\frac{d\omega}{\pi}\,
\big[n_{\rm B}(\omega)+n_{\rm F}(\omega+\varepsilon)\big]
\nonumber\\
&\times
\sum_{\nu,\nu',\lambda'}
\int\frac{d^2{\bm q}}{(2\pi)^2}\,
\mathrm{Im} D^{\nu\nu'}({\bm q},\omega)\,
\mathrm{Im} G^{(0)}_{\lambda'}({\bm k}+{\bm q},\varepsilon+\omega)
\nonumber\\
&\times
j^{\nu}_{{\bm k},\lambda;{\bm k}+{\bm q},\lambda'}
j^{\nu'}_{{\bm k}+{\bm q},\lambda';{\bm k},\lambda}.
\label{eq:app_sigma_ret_im}
\end{align}

We now take the low-energy limit at the Fermi surface. Evaluating the bare
retarded Green's function,
$G^{R,(0)}_{\lambda'}({\bm p},\epsilon)
=1/(\epsilon-\xi_{{\bm p},\lambda'}+i0^+)$, gives
\begin{equation}
\mathrm{Im} G^{R,(0)}_{\lambda'}({\bm k}+{\bm q},\varepsilon+\omega)
=
-\pi\,\delta\big(\varepsilon+\omega-\xi_{{\bm k}+{\bm q},\lambda'}\big),
\label{eq:app_ImG_delta}
\end{equation}
which pins the intermediate fermionic state to the mass shell
$\xi_{{\bm k}+{\bm q},\lambda'}=\varepsilon+\omega$.

In the $T\to0$ limit, the thermal factor in
Eq.~\eqref{eq:app_sigma_ret_im} simplifies to
\begin{equation}
n_{\rm B}(\omega)+n_{\rm F}(\omega+\varepsilon)
=
-\Theta(-\omega)+\Theta(-\omega-\varepsilon).
\label{eq:app_T0_factor}
\end{equation}
For \(\varepsilon>0\), this equals \(-1\) for \(\omega\in(-\varepsilon,0)\)
and vanishes otherwise, while for \(\varepsilon<0\) it equals \(+1\) for
\(\omega\in(0,-\varepsilon)\). This yields the compact frequency window
\(\omega\in[\min(0,-\varepsilon),\max(0,-\varepsilon)]\).

Finally, for \({\bm k}\simeq{\bm k}_{\rm F}\), the on-shell condition in
Eq.~\eqref{eq:app_ImG_delta} implies that
\({\bm k}'={\bm k}+{\bm q}\) also lies close to the Fermi surface. In this
limit, the longitudinal and transverse current matrix elements can be
constructed from Eq.~\eqref{eq:app_jxy} as
\begin{align}
j^{\rm L}_{{\bm k},\lambda;{\bm k}',\lambda'}
&=
\hat{\bm q}\cdot{\bm j}_{{\bm k},\lambda;{\bm k}',\lambda'}
=v_{\rm F}\frac{i(\lambda'-\lambda)}{2}\,
e^{i(\varphi_{{\bm k}'}-\varphi_{\bm k})/2}\nonumber\\
&\hspace{2.5cm}\times\mathrm{sign}\Big(\sin{\dfrac{\varphi_{\bm k}-\varphi_{{\bm k}'}}{2}}\Big),
\\
j^{\rm T}_{{\bm k},\lambda;{\bm k}',\lambda'}
&=
(\hat{\bm z}\times\hat{\bm q})\cdot
{\bm j}_{{\bm k},\lambda;{\bm k}',\lambda'}
=v_{\rm F}\frac{\lambda'+\lambda}{2}\,
e^{i(\varphi_{{\bm k}'}-\varphi_{\bm k})/2}
\nonumber\\&\hspace{2.8cm}\times\mathrm{sign}\Big(\sin{\dfrac{\varphi_{\bm k}-\varphi_{{\bm k}'}}{2}}\Big).
\label{eq:app_jLT}
\end{align}
At the Fermi surface one has \(\lambda'=\lambda\), so the longitudinal
matrix element vanishes and only the transverse channel survives. Eq.~
\eqref{eq:app_sigma_ret_im} therefore reduces to
\begin{align}
&\mathrm{Im}\Sigma^{(R)}_{+}({\bm k},\varepsilon)
=
- v_{\rm F}^2\;\mathrm{sign}(-\varepsilon)
\int_{\min(0,-\varepsilon)}^{\max(0,-\varepsilon)}\frac{d\omega}{\pi}
\int\frac{d^2{\bm q}}{(2\pi)^2}
\nonumber\\
&\times
\mathrm{Im} D^{TT}({\bm q},\omega)\,
\mathrm{Im} G^{(0)}_{+}({\bm k}+{\bm q},\varepsilon+\omega),
\label{eq:Self_energy_final_app_B}
\end{align}
which is the transverse-only form of the self energy.
%========================================================
\section{Bare and RPA-dressed transverse phonon propagator}
\label{app:rpa_phonon}

In this appendix we summarize the steps to obtain the dressed
transverse phonon propagator and the dimensionless self-energy used in the
main text. We first derive the bare transverse-transverse (TT) phonon propagator, then discuss its RPA-like dressing in the
transverse current channel, and finally show the angular integration used
to obtain the reduced self-energy.

The components of the vector potential are linear combinations of phonon coordinates, as seen in Eq.~(\ref{eq:gaugephonons}).
The phonon propagator is canonically defined as
\begin{align}
D^{\nu\nu'}({\bm q},i\omega_m)
=
-\int_0^\beta d\tau\, e^{i\omega_m\tau}
\left\langle {\cal T}_\tau\, A_{{\bm q},\nu}(\tau)\,A_{-{\bm q},\nu'}(0)\right\rangle.
\label{eq:app_D_def}
\end{align}
Using the free-phonon correlator
\begin{align}
&-\int_0^\beta d\tau\, e^{i\omega_m\tau}
\Big\langle {\cal T}_\tau\,
\big(a_{{\bm q},\nu}(\tau)+a^\dagger_{-{\bm q},\nu}(\tau)\big)
\nonumber\\
&\times
\big(a_{-{\bm q},\nu}(0)+a^\dagger_{{\bm q},\nu}(0)\big)
\Big\rangle
=
\frac{2\omega_{{\bm q},\nu}}{(i\omega_m)^2-\omega_{{\bm q},\nu}^2},
\label{eq:app_freephonon_corr}
\end{align}
and one finds for the bare TT component
\begin{align}
&D^{(0),\rm{TT}}({\bm q},i\omega_m)
=
\sum_{\nu={\rm L,T}}
g_\nu^2(q)\ \nonumber \\
&\times\big[M_{{\rm T},\nu}(\hat{\bm q})\big]^2
\frac{2\omega_{{\bm q},\nu}}{(i\omega_m)^2-\omega_{{\bm q},\nu}^2}.
\label{eq:app_D0_TT_matsubara}
\end{align}
After analytical continuation, 
\begin{align}
&D^{(0),\rm{TT}}({\bm q},\omega) 
= g_{L}^2(q) \sin^2(\varphi_{{\bm q}}) \frac{2\omega_{{\bm q},L}}{(\omega+i0^+)^2- \omega_{{\bm q},L}^2} \nonumber \\
&+ g_{\rm{T}}^2(q) \cos^2(\varphi_{{\bm q}}) \frac{2\omega_{{\bm q},\rm{T}}}{(\omega+i0^+)^2- \omega_{{\bm q},\rm{T}}^2}~.
\label{eq:app_D0_TT_expression_sum}
\end{align}

We use the approximation of a single effective mode of frequency $\omega_{\bm q} \approx \omega_{{\bm q},{\rm L}}\approx \omega_{{\bm q},{\rm T}}$, which leads to 

\begin{align}
  D^{(0), TT}({\bm q},\omega) \simeq { g}^2(q) \frac{2{ \omega}_{{\bm q}}}{(\omega+i0^+)^2 - { \omega}_{{\bm q}}^2}~,
\label{eq:app_D0_TT_expression}
\end{align}
where, $g(q) = (g_2/v_{\rm F}) \times \sqrt{q^2/(2 \rho_m { \omega}_{{\bm q}})}$. Following Eq.~(\ref{eq:chiT_expand}) and the Dyson equation,
$
\big[D^{\rm{TT}}({\bm q},\omega)\big]^{-1}
=
\big[D^{(0),\rm{TT}}({\bm q},\omega)\big]^{-1}
-
\chi^{(\rm{T})}({\bm q},\omega),
$
 we calculate the dressed transverse phonons propagator. In the dressed propagator, the Landau-damping term provides the relevant imaginary part of the denominator (with $\alpha>0$). We therefore suppress the explicit infinitesimal \(i0^+\) in what follows and write

\begin{align}
    D^{\rm{TT}}({\bm q},\omega)& = \left[ \frac{\omega^2 - {\ \omega}_{{\bm q}}^2}{2g^{2}(q){ \omega}_{{\bm q}}} + 
\frac{4\pi \chi_{\rm orb}}{e^2\mu_0} q^2
+
i\,\alpha\frac{\omega}{v_{\rm F}q}\right]^{-1}
\nn
&= \frac{2g^{2}(q)\omega_{\bm{q}}}{ \displaystyle \omega^2 - { \omega}_{{\bm q}}^2 + \frac{g_2^2}{v_{\rm F}^2\rho_m}\left(
\frac{4\pi \chi_{\rm orb}}{e^2\mu_0} q^4
+
i\,\alpha\frac{\omega}{v_{\rm F}q}\right) }
\nn
&=
g^2(q) \frac{2 { \omega}_{{\bm q}}}{{\omega^2-\omega_{\bm q}^2 + v_{\mathrm{F}}^2\bar{\chi}_{\text{orb}}q^4/k_{\mathrm{F}}^2+i\bar\alpha v_{\mathrm{F}} q\omega}}
~.
\label{eq:Transverse_phonon_propgator}
\end{align}

We now show the angular integration used after inserting the dressed
propagator into the low-energy self-energy. Defining
\begin{align}
&{\cal I}(k,q,\varepsilon+\omega+\mu)
\equiv
\int_0^{2\pi}\frac{d\varphi_{\bm q}}{2\pi}\,
\mathrm{Im}\,G^{(0)}_{+}({\bm k}+{\bm q},\varepsilon+\omega+i0^+)
\nonumber\\
&=
-\dfrac{1}{2}\int_0^{2\pi}d\varphi_{\bm q}\,
\delta\!\left(\varepsilon+\omega+\mu-v_{\rm F}|{\bm k}+{\bm q}|\right),
\label{eq:app_I_def}
\end{align}
the delta function implies $\varepsilon+\omega+\mu>0$ and, choosing
$\varphi_{\bm k}=0$,
\begin{align}
\cos\varphi_{\bm q}
=
\frac{(\varepsilon+\omega+\mu)^2-v_{\rm F}^2(k^2+q^2)}
{2v_{\rm F}^2kq}.
\label{eq:app_cos_phi}
\end{align}
The condition $-1\le\cos\varphi_{\bm q}\le1$ restricts the momentum
integration to $q_{\rm min}<q<q_{\rm max}$, with
\begin{align}
v_{\rm F}q_{\rm min}
&=|\varepsilon+\omega+\mu-v_{\mathrm{F}}k|,
\nonumber\\
v_{\rm F}q_{\rm max}
&=
\varepsilon+\omega+\mu+v_{\rm F}k.
\label{eq:app_q_bounds}
\end{align}
Under these conditions,
\begin{align}
&{\cal I}(k,q,\varepsilon+\omega+\mu)
=
-\,2\,|\varepsilon+\omega+\mu|\Theta(\varepsilon+\omega+\mu)
\nonumber\\
&\times
\Big[(2v_{\rm F}^2kq)^2
-\big((\varepsilon+\omega+\mu)^2-v_{\rm F}^2(k^2+q^2)\big)^2\Big]^{-1/2}.
\label{eq:app_I_final}
\end{align}
Substituting the results from Eq.~(\ref{eq:Transverse_phonon_propgator}) and (\ref{eq:app_I_final}) into the low-energy self-energy in Eq.~(\ref{eq:Self_energy_final_app_B}), we find that
\begin{align}
&\mathrm{Im}\Sigma_+^{(R)}({\bm k},\varepsilon) = v_{\rm F}^2 \mathrm{sign}(-\varepsilon)\int_{\omega_{\rm min}}^{\omega_{\rm max}} \frac{d\omega}{\pi} \int_{q_{\rm min}}^{q_{\rm max}} \frac{d q}{\pi}
\nonumber \\ &\times\frac{q |\varepsilon + \omega+ \mu|\Theta(\varepsilon+\omega+\mu)}{\sqrt{(2 v_{\rm F}^2 k q)^2 - \big[(\varepsilon + \omega + \mu)^2 - v_{\rm F}^2(k^2+q^2)\big]^2}}
\nonumber \\
&\times
g^2(q) ~\mathrm{Im} \left[ \frac{2 { \omega}_{\bm q}}{ \displaystyle \omega^2 - { \omega}_{{\bm q}}^2 + v_{\mathrm{F}}^2{\bar \chi_{\rm orb}} q^4/k_{\mathrm{F}}^2 + i v_{\mathrm{F}} { \bar{\alpha}} q \omega }\right]
~,
\label{eq:dimmensional_self_energy_app_C}
\end{align}

\begin{align}
&\mathrm{Im}\Sigma_+^{(R)}(k_{\mathrm{F}},\varepsilon>0) = -\frac{g_{\rm 2}^2}{\rho_{m}} \int_{-\varepsilon}^{0} \frac{d\omega}{\pi} \int_{(\varepsilon+\omega)/{v_{\mathrm{F}}}}^{2k_{\rm F}+(\varepsilon+\omega)/v_{\mathrm{F}}} \frac{d q}{\pi}
\nonumber \\ &\times\frac{q |\varepsilon + \omega+ \mu|\Theta(\varepsilon+\omega+\mu)}{\sqrt{(2 v_{\rm F}^2 k_{\mathrm{F}} q)^2 - \big[(\varepsilon + \omega + \mu)^2 - v_{\rm F}^2(k_{\mathrm{F}}^2+q^2)\big]^2}}
\nonumber \\
&\times
 ~\  \frac{\bar{\alpha}v_{\mathrm{F}} q^3|\omega|}{ \displaystyle (\omega^2 - (c_{\rm ph}q)^2 -  v_{\mathrm{F}}^2\bar{\chi}_{\rm orb} q^4/k_{\mathrm{F}}^2 )^2 +  (\bar{\alpha} v_{\mathrm{F}} q \omega)^2 }
~.
\label{eq:dimmensional_self_energy_app_C}
\end{align}
Rescaling Eq.~(\ref{eq:dimmensional_self_energy_app_C}) to dimensionless variables leads to
Eq.~\eqref{eq:sigma_analit_final_dimmensionless} in the main text.

%========================================================
\section{Power laws of the self-energy}
\label{app:powerlaws}

In this appendix we extract the low-energy power laws of the imaginary part
of the retarded self-energy at the Fermi surface from the low-energy form of
the dressed TT phonon propagator. Throughout we set
$k\simeq k_{\rm F}$ and consider
$0<\bar\varepsilon\ll 1$, so that at $T=0$ the frequency transfer is
restricted to $\bar\omega\in[-\bar\varepsilon,0]$. It is convenient to write the
self-energy in the schematic form
\begin{align}
\mathrm{Im} \bar\Sigma^{R}_{+}(1,\bar\varepsilon)
&\simeq
-C
\int_{-\bar \varepsilon}^{0}\frac{d\bar\omega}{\pi}
\int_{\bar \varepsilon+\bar \omega}^{2+\bar \varepsilon+\bar \omega}\frac{d\bar q}{\pi}\,
{\cal K}(\bar q, \bar \omega, \bar \varepsilon)
\nonumber\\
&\times
\frac{\bar\alpha\,\bar q^3|\bar \omega|}
{\big[\bar\omega^2-(c_{\rm ph}/v_{\rm F})^2 \bar q^2 +\bar\chi_{\rm orb} \bar q^4\big]^2
+\bar\alpha^{2} \bar q^2 \bar\omega^2},
\label{eq:PL_start}
\end{align}
where $C=g_2^2k_{\mathrm{F}}/(\rho_m v_{\mathrm{F}}^3)$ absorbs smooth prefactors and factor ${\cal K}$ is defined as
\begin{equation}
    {\cal K}=\dfrac{\bar{q}(\bar{\varepsilon}+\bar{\omega}+1)\Theta(\bar{\varepsilon}+\bar{\omega}+1)}{\sqrt{(2\bar{q})^2-((\bar{\varepsilon}+\bar{\omega}+1)^2-(1+\bar{q}^2))^2}}.
\end{equation}
This is smooth and finite away from the integration endpoints. At the endpoints, ${\cal K}$ presents integrable singularities. Thus, the scaling is governed by the bosonic kernel.

It is useful to keep in mind that the frequency integral is dominated by
$|\bar \omega|\sim \bar\varepsilon$. The scaling of the self-energy is therefore
determined by the characteristic momenta selected by the peak of the bosonic
factor. In the large $\bar\alpha$ regime the dressed phonon is overdamped,
and this is precisely the regime in which anomalous scaling emerges.
However, overdamping alone does not guarantee non-Fermi-liquid behavior.
Depending on which momentum region dominates the integral, the self-energy
may still remain Fermi-liquid-like.

\subsection{Overdamped reduction for large $\bar\alpha$}
When $\bar\alpha$ is large, the characteristic momenta selected by the
bosonic kernel satisfy $\bar q\gg|\bar \omega|$ in the regimes of interest. One may
then neglect the $\bar \omega^2$ term inside the square brackets of
Eq.~\eqref{eq:PL_start}, so that
\begin{align}
&\big[\bar \omega^2-(c_{\rm ph}/v_{\rm F})^2 \bar q^2 + \bar\chi_{\rm orb} \bar q^4\big]^2
+\bar\alpha^{\,2}\bar q^2 \bar \omega^2
\nonumber\\
&\qquad\simeq
\bar{q}^2\Big(\bar q^2\Big[(c_{\rm ph}/v_{\rm F})^2- \bar\chi_{\rm orb} \bar q^2\Big]^2
+\bar\alpha^{\,2}\bar\omega^2\Big).
\label{eq:PL_overdamped_den}
\end{align}
The bosonic factor then reduces to
\begin{align}
&\frac{\bar\alpha\,\bar q^3|\bar\omega|}
{\big[\bar\omega^2-(c_{\rm ph}/v_{\rm F})^2\bar q^2+\bar\chi_{\rm orb} \bar q^4\big]^2
+\bar\alpha^{\,2}\bar q^2 \bar\omega^2} \nonumber\\
&\simeq
\frac{\bar\alpha\,\bar q|\bar\omega|}
{\bar q^2\big[(c_{\rm ph}/v_{\rm F})^2 - \bar\chi_{\rm orb}\bar q^2\big]^2
+\bar\alpha^{\,2}\bar\omega^2}.
\label{eq:PL_overdamped_kernel}
\end{align}
The scaling of the self-energy is therefore controlled by the momenta at
which the denominator of Eq.~\eqref{eq:PL_overdamped_kernel} is minimal.

\subsection{Diamagnetic case: Fermi-liquid regime}

We first consider the regime in which the quartic term is negligible,
$\textit{i.e.}$ when $|\bar\chi_{\rm orb}| \bar q^2\ll (c_{\rm ph}/v_{\rm F})^2$. In this case
the denominator of Eq.~\eqref{eq:PL_overdamped_kernel} reduces to
\begin{equation}
(c_{\rm ph}/v_{\rm F})^4 \bar q^2+\bar\alpha^{\,2} \bar \omega^2,
\end{equation}
so the integrand is peaked at
\begin{align}
q_{\rm FL}
\sim
\frac{\bar\alpha}{(c_{\rm ph}/v_{\rm F})^2}\,|\bar \omega|.
\label{eq:q_first_app}
\end{align}
Since the frequency integral is restricted to $|\bar\omega|\lesssim \bar \varepsilon$,
the dominant contribution comes from $|\bar\omega|\sim\bar\varepsilon$, giving
\begin{equation}
q_{\rm FL}\sim
\frac{\bar\alpha}{(c_{\rm ph}/v_{\rm F})^2}\,\bar\varepsilon.
\end{equation}
Thus $\bar q$ scales linearly with $\bar \varepsilon$. We may then rescale the integration variables as
$\bar\omega={\tilde\omega}\,\bar\varepsilon$ and $\bar{q}=\tilde{q}\,\bar\varepsilon$ in
Eq.~\eqref{eq:PL_start}. Since the measure contributes
$d\bar\omega\,d \bar q\sim \bar\varepsilon^2$, while both ${\cal K}$ and the overdamped
kernel remain finite in this scaling limit, one obtains
\begin{align}
&\mathrm{Im} \bar\Sigma_+^{(R)}(1, \bar\varepsilon)
\simeq
-\frac{g_2^2k_{\mathrm{F}}}{\rho_m v_{\mathrm{F}}^3}
\int_{-1}^{0}\frac{d\tilde{\omega}}{\pi}\,\bar\varepsilon
\int_{1+\tilde{\omega}}^{2/\bar\varepsilon+(1+\tilde{\omega})}
\frac{d\tilde{q}}{\pi}\,\bar\varepsilon
\nonumber\\
&\times
\frac{\bar\varepsilon^2 \tilde{q}\,(1+\tilde{\omega}+1/\bar\varepsilon)}
{\bar \varepsilon^2
\sqrt{
4\tilde{q}^2/\bar \varepsilon^2
-
\Big[(1+\tilde{\omega}+1/\bar \varepsilon)^2-1/\bar \varepsilon^2-\tilde{q}^2\Big]^2
}}
\nonumber\\
&\times
\frac{\bar\varepsilon^2 \bar\alpha\,\tilde{q}|\tilde{\omega}|}
{\bar\varepsilon^2\Big((c_{\rm ph}/v_{\rm F})^4\tilde{q}^2+\bar\alpha^2\tilde{\omega}^2\Big)}
\nonumber\\
&\to
-\bar\varepsilon^2\frac{g_2^2k_{\mathrm{F}}}{\rho_m v_{\mathrm{F}}^3}
\int_{-1}^{0}\frac{d\tilde{\omega}}{\pi}
\int_{1+\tilde{\omega}}^{\infty}\frac{d\tilde{q}}{\pi}
\nonumber\\
&\times
\frac{\tilde{q}}{2\sqrt{\tilde{q}^2-(1+\tilde{\omega})^2}}
\frac{\bar\alpha\,\tilde{q}|\tilde{\omega}|}
{(c_{\rm ph}/v_{\rm F})^4\tilde{q}^2+\bar\alpha^2\tilde{\omega}^2}.
\label{eq:PL_FL_rescaled_limit}
\end{align}
Hence the self-energy remains Fermi-liquid like, even though the phonon mode
is overdamped.

\subsection{Diamagnetic case: non-Fermi-liquid regime}

We now consider the regime in which the quartic term dominates over the
bare phonon dispersion,
\begin{equation}
(c_{\rm ph}/v_{\rm F})^2 \ll |\bar\chi_{\rm orb}| \bar{q}^2.
\end{equation}
Then the $\bar q$-dependent part of the denominator in
Eq.~\eqref{eq:PL_overdamped_kernel} becomes
$\sim \bar\chi_{\rm orb}^{2}\bar q^6$, and the bosonic kernel is peaked at
\begin{align}
q_{\rm NFL}
\sim
\left(
\frac{\bar\alpha|\bar\omega|}{|\bar\chi_{\rm orb}|}
\right)^{1/3}.
\label{eq:q_second_app}
\end{align}
Again taking $|\bar\omega|\sim \bar\varepsilon$, this gives
\begin{equation}
q_{\rm NFL}
\sim
\left(
\frac{\bar\alpha \bar\varepsilon}{|\bar\chi_{\rm orb}|}
\right)^{1/3}.
\end{equation}
In this regime $q_{\rm NFL}\gg|\bar\omega|$ parametrically as
$\bar\varepsilon\to0$, so the overdamped approximation is consistent.
We now scale
\begin{equation}
\bar\omega=\tilde{\omega}\,\bar \varepsilon,
\qquad
\bar q=\tilde{q} (\bar \varepsilon)^{1/3}.
\end{equation}
Then Eq.~\eqref{eq:PL_start} results in,
\begin{align}
&\mathrm{Im} \bar\Sigma_+^{(R)}(1,\bar\varepsilon)
\to
-\bar\varepsilon^{2/3}\frac{g_2^2k_{\mathrm{F}}}{\rho_m v_{\mathrm{F}}^3}
\int_{-1}^{0}\frac{d\tilde{\omega}}{\pi}
\int_{0}^{\infty}\frac{d\tilde{q}}{2\pi} \\
&\times
\frac{\bar\varepsilon^{4/3} \tilde{q}\,|1+\tilde{\omega}+1/\bar\varepsilon|}
{\bar \varepsilon^{4/3}
\sqrt{
4\tilde{q}^2/\bar \varepsilon^2
-
\Big[(1+\tilde{\omega}+1/\bar \varepsilon)^2-1/\bar \varepsilon^2-\tilde{q}^2\Big]^2
}}
\nonumber\\
&\times \frac{\bar\alpha\,\tilde{q}\tilde{\omega}}
{\bar\chi_{\rm orb}^2\tilde{q}^6+\bar\alpha^2\tilde{\omega}^2}.
\label{eq:PL_NFL_rescaled}
\end{align}
The remaining integral is finite, so that
\begin{align}
\mathrm{Im} \bar\Sigma^{R}_{+}(1,\bar\varepsilon)
\propto \bar\varepsilon^{2/3}.
\label{eq:PL_NFL}
\end{align}

\subsection{Diamagnetic case: crossover scale}
\label{app:crossover_diamagnetic}

The crossover between the Fermi-liquid like and non-Fermi-liquid regimes
occurs when the characteristic momenta
Eqs.~\eqref{eq:q_first_app} and \eqref{eq:q_second_app} become comparable:
\begin{align}
\frac{\bar\alpha}{(c_{\rm ph}/v_{\rm F})^2}\,\bar\varepsilon
\sim
\left(
\frac{\bar\alpha\bar\varepsilon}{|\bar\chi_{\rm orb}|}
\right)^{1/3}.
\end{align}
Solving for $\bar\varepsilon$ gives
\begin{align}
\varepsilon_\ast
\sim
\frac{(c_{\rm ph}/v_{\rm F})^3}
{\bar\alpha\sqrt{|\bar\chi_{\rm orb}|}}.
\label{eq:eps_star_app}
\end{align}
For $\varepsilon\ll\varepsilon_\ast$ the self-energy is Fermi-liquid like,
Eq.~\eqref{eq:PL_FL_rescaled_limit}, whereas for $\varepsilon\gg\varepsilon_\ast$ it crosses
over to the non-Fermi-liquid form Eq.~\eqref{eq:PL_NFL}.

\subsection{Paramagnetic case: marginal-Fermi-liquid regime}

\label{app:marginal_fermi_liquid}

We now consider the paramagnetic case $\bar\chi_{\rm orb}>0$. The real part of the bosonic denominator softens at the finite momentum
\begin{align}
q_{\rm shell} =
\frac{(c_{\rm ph}/v_{\rm F})}{\sqrt{|\bar\chi_{\rm orb}}|}.
\label{eq:q0_soft_app}
\end{align}
This is the soft shell discussed in the main text.

In the overdamped limit, the bosonic denominator becomes
\begin{equation}
\bar q^2\Big[(c_{\rm ph}/v_{\rm F})^2-\bar\chi_{\rm orb} \bar q^2\Big]^2
+\bar\alpha^{\,2}\bar \omega^2,
\end{equation}
which is minimized near $\bar q=q_{\rm{shell}}$. We therefore write
\begin{equation}
\bar q=q_{\rm shell}+\delta \bar q,
\qquad
|\delta \bar q|\ll q_{\rm{shell}}.
\end{equation}
Expanding the real part of the denominator around $q_{\rm{shell}}$ gives
\begin{align}
(c_{\rm ph}/v_{\rm F})^2-\bar\chi_{\rm orb}\bar q^2
&=
(c_{\rm ph}/v_{\rm F})^2-\bar\chi_{\rm orb}(q_{\rm{shell}}+\delta \bar q)^2
\nonumber\\
&\simeq
-2\bar\chi_{\rm orb}q_{\rm{shell}}\,\delta \bar q,
\end{align}
since $(c_{\rm ph}/v_{\rm F})^2-\bar\chi_{\rm orb}q_{\rm{shell}}^2=0$ by definition of
$q_{\rm{shell}}$. Substituting this into the denominator yields
\begin{align}
&q^2\Big[(c_{\rm ph}/v_{\rm F})^2-\bar\chi_{\rm orb}q^2\Big]^2
+\bar\alpha^{2}\omega^2
\nonumber\\
&\qquad\simeq
4{\bar \chi}_{\rm orb}^{2}q_{\rm{shell}}^4(\delta q)^2+\bar\alpha^{2}\omega^2.
\label{eq:soft_den_app}
\end{align}
Thus, close to the shell, the bosonic factor takes a Lorentzian form in
$\delta q$,
\begin{align}
\frac{\bar\alpha \bar q|\bar\omega|}
{\bar q^2\big[(c_{\rm ph}/v_{\rm F})^2-\bar\chi_{\rm orb}\bar q^2\big]^2
+\bar\alpha^{\,2}\bar\omega^2}
\simeq
\frac{\bar\alpha q_{\rm{shell}}|\bar \omega|}
{4\bar\chi_{\rm orb}^{2}q_{\rm{shell}}^4(\delta \bar q)^2+\bar\alpha^{2}\bar\omega^2}.
\label{eq:soft_shell_lorentz}
\end{align}
The apparent singularity at $q=q_{\rm{shell}}$ is therefore regulated by the damping
term and does not produce a divergence.
\par
Balancing the two terms in Eq.~\eqref{eq:soft_den_app} gives the shell width
\begin{align}
\delta q_{\rm shell}
\sim
\frac{\alpha}{2\bar\chi_{\rm orb}q_{\rm{shell}}^2}\,|\bar \omega|
=
\frac{\alpha}{2(c_{\rm ph}/v_{\rm F})^{2}}\,|\bar\omega|.
\label{eq:soft_width_app}
\end{align}
Since the remaining kinematic factors are smooth at finite momentum
$\bar q\simeq q_{\rm shell}$, the singular part of the $\bar q$ integral reduces to
\begin{align}
I_{\rm shell}(\omega)
&\sim
\int_{-\delta q_{\rm shell}/2}^{+\delta q_{\rm shell}/2}
d(\delta \bar q)\,
\frac{\bar\alpha q_{\rm shell}|\omega|}
{4\bar\chi_{\rm orb}^{2}q_{\rm shell}^4(\delta q)^2+\bar\alpha^{2}\omega^2},
\label{eq:Ishell_deltaq}
\end{align}
We now rescale
\begin{equation}
\delta \bar q=|\bar \omega|x,
\qquad
d(\delta \bar q)=|\bar \omega|dx.
\end{equation}
Then Eq.~\eqref{eq:Ishell_deltaq} becomes
\begin{align}
I_{\rm shell}(\omega)
&\sim
\int_{x_{\rm min}}^{x_{\rm max}} |\omega|\,dx\,
\frac{\bar\alpha q_{\rm shell}|\omega|}
{4\bar\chi_{\rm orb}^{2}q_{\rm shell}^4\omega^2x^2+\bar\alpha^{2}\omega^2}
\nonumber\\
&\sim
\int_{x_{\rm min}}^{x_{\rm max}} dx\,
\frac{\bar\alpha q_{\rm shell}}
{4\bar\chi_{\rm orb}^{2}q_{\rm shell}^4x^2+\bar\alpha^{2}},
\label{eq:Ishell_x}
\end{align}
with
\begin{equation}
x_{\rm min}=-\frac{\alpha}{4(c_{\rm ph}/v_{\rm F})^{2}}\,,
\qquad
x_{\rm max}=\frac{\alpha}{4(c_{\rm ph}/v_{\rm F})^{2}}\,.
\end{equation}
The shell momentum is always in the domain of integration $\textit{i.e.}$, $q_{\rm min} <  q_{\rm shell} < q_{\rm max}$, unless $\bar \chi_{\rm orb} \rightarrow0$ and thus the argument here holds for a broad range of parameters ($c_{\rm ph}, v_{\rm F} \,\rm {and}\, |\bar{\chi}_{\rm orb}|$). As the Lorentzian is independent of $\omega$ up to a finite dimensionless constant.
We have,
\begin{equation}
I_{\rm shell}(\omega)\sim \omega^0.
\end{equation}
Since the $\omega$ integral extends over a window of width $\varepsilon$, the
soft-shell contribution to the self-energy scales as
\begin{align}
\mathrm{Im}\Sigma_+^{(R)}(k_{\rm F},\varepsilon)
\propto \varepsilon.
\label{eq:PL_MFL}
\end{align}
Hence, the paramagnetic regime $\bar\chi_{\rm orb}>0$ yields a marginal-Fermi-liquid-like contribution
dominated by the soft shell. This holds for arbitrarily low $|\omega|$ and in the limit $\omega \rightarrow0$, the contribution from the Lorentzian is parametrically larger than the Fermi-liquid behavior seen in Eq. (\ref{eq:PL_FL_rescaled_limit}).

\subsection{Paramagnetic case: crossover scale}
\label{app:crossover_paramagnetic}

At higher energies the large-$q$ part of the overdamped kernel can again
take over, leading to the same crossover scale as in the diamagnetic case ($\bar\chi_{\rm orb}<0$). As the non-Fermi liquid scaling is parametrically larger than the marginal-Fermi-liquid scaling, a crossover occurs. Equating the soft-shell momentum $q_{\rm{shell}}$ with the non-Fermi-liquid momentum scale gives
\begin{equation}
\frac{(c_{\rm ph}/v_{\rm F})}{\sqrt{|\bar\chi_{\rm orb}|}}
\sim
\left(
\frac{\bar\alpha\bar\varepsilon}{{|\bar\chi_{\rm orb}|}}
\right)^{1/3},
\end{equation}
and therefore
\begin{equation} 
\varepsilon_\ast
\sim
\frac{(c_{\rm ph}/v_{\rm F})^3}{
{\bar\alpha\sqrt{|\bar\chi_{\rm orb}|}}}.
\end{equation}
Thus, the paramagnetic case exhibits a marginal-Fermi-liquid infrared
behavior, followed by a crossover into the same non-Fermi-liquid regime
encountered in the diamagnetic case. Notably, the energy scale at which this crossover happens is the same in both cases.

\subsection{Weakly damped estimate of the self energy }
\label{app:small_alpha}

For weak to moderate damping, the overall magnitude of the self-energy is controlled by different momentum regions in the diamagnetic and paramagnetic regimes. We give a simple scaling estimates for the magnitude of the self-energy relevant to Fig \ref{fig:3}.

In the diamagnetic case, and for sufficiently small $|\bar\chi_{\rm orb}|$, the bosonic kernel in Eq.~\eqref{eq:PL_start} is sharply peaked at the solution of
\begin{equation}
\bar\omega^2-\left(c_{\rm ph}/v_{\rm F}\right)^2\bar q^2-\bar\chi_{\rm orb}\bar q^4=0.
\end{equation}
Neglecting the quartic term, one has $\bar q_\ast \simeq (|\bar\omega|v_{\rm F})/({c}_{\rm ph})$, and the Lorentzian may be reduced to a Dirac delta function. This gives
\begin{equation}
\frac{\bar\alpha\,\bar q^3|\bar\omega|}
{\big[\bar\omega^2-(c_{\rm ph}/v_{\rm F})^2\bar q^2\big]^2+\bar\alpha^2\bar q^2\bar\omega^2}
\;\to\;
\frac{\pi}{2}\,
\frac{|\bar\omega|}{(c_{\rm ph}/v_{\rm F})^3}\,
\delta\!\left(\bar q-\bar q_\ast\right),
\end{equation}
so that
\begin{equation}
\mathrm{Im}\,\bar\Sigma^{R}_{+}(1,\bar\varepsilon)
\sim
\frac{\bar\varepsilon^2}{(c_{\rm ph}/v_{\rm F})^3}.
\label{eq:smallalpha_dia}
\end{equation}

By contrast, in the paramagnetic case $\bar\chi_{\rm orb}>0$, the dominant contribution is controlled by the softened shell
\begin{equation}
 q_{\rm shell}=\frac{(c_{\rm ph}/v_{\rm F})}{\sqrt{\bar\chi_{\rm orb}}}.
\end{equation}. The shell contribution is estimated as before,
\begin{equation}
I_{\rm shell}(\bar\omega)\sim
\int d(\delta\bar q)\,
\frac{\bar\alpha\, q_{\rm shell}^3|\bar\omega|}
{\left[\bar\omega^2+2(c_{\rm ph}/v_{\rm F})^2 q_{\rm shell}\delta\bar q\right]^2
+\left(\bar\alpha q_{\rm shell}\bar\omega\right)^2}.
\end{equation}
Integrating over $\delta \bar q$, one finds,
\begin{equation}
I_{\rm shell}(\bar\omega)
\sim
\frac{\pi}{2}\frac{\bar q_{\rm shell}}{(c_{\rm ph}/v_{\rm F})^2}
\sim
\frac{1}{(c_{\rm ph}/v_{\rm F})\sqrt{\bar\chi_{\rm orb}}},
\end{equation}
and therefore
\begin{equation}
\mathrm{Im}\,\bar\Sigma^{R}_{+}(1,\bar\varepsilon)
\sim
\frac{\bar\varepsilon}{(c_{\rm ph}/v_{\rm F})\sqrt{\bar\chi_{\rm orb}}}.
\label{eq:smallalpha_para}
\end{equation}

Equations~\eqref{eq:smallalpha_dia} and \eqref{eq:smallalpha_para} show that the diamagnetic and paramagnetic regimes are controlled by different momentum structures of the bosonic kernel. In the former, the magnitude is set by the sharp Dirac-delta reduction and carries the factor $(c_{\rm ph}/v_{\rm F})^{-3}$. In the latter, the dominant contribution comes from the broadened soft shell, yielding instead the weaker dependence $[(c_{\rm ph}/v_{\rm F})\sqrt{\bar\chi_{\rm orb}}]^{-1}$. This helps in understanding and estimating the self energy scaling seen in Fig \ref{fig:3} (a) and (b).

\section{Numerical estimates for $\bar{\chi}_{\rm orb}$ and $\bar{\alpha}$}
\label{app:parameters}
In this appendix, we estimate the magnitude of the orbital response $\bar{\chi}_{\rm orb}$ and the damping coefficient $\bar{\alpha}$ for graphene and MATBG used in the main text.

In the diamagnetic case, we first evaluate the magnetic susceptibility $\chi_m$ using Ref.~\cite{gomez2011measurable}. Using the lattice estimate $|\chi_m| \sim \mu_0 e^2 |t| a^2/\hbar^2$, where $t$ is the nearest-neighbor hopping amplitude ($2.7~{\rm eV}$) and $a$ is the lattice constant ($2.86~\text{\AA}$), together with the relation $\chi_m = (\mu_0 e^2/4\pi)\,\chi_{\rm orb}$, we obtain $\chi_{\rm orb} = -9.8 \times 10^{30}\ {\rm m}^2\,{\rm J}^{-1}\,{\rm s}^{-2}$ in SI units. We therefore take $\chi_{\rm orb} = -9.8 \times 10^{30}\ {\rm m}^2\,{\rm J}^{-1}\,{\rm s}^{-2}$ as a representative diamagnetic value. To estimate $\alpha$, we use the imaginary part of the leading-order transverse current-current response of doped graphene. As reported in Ref.~\cite{principi2009linear}, $\mathrm{Im}\,\chi^{(T)}(q,\omega) = (N_f \varepsilon_F)/(2\pi\hbar^2)\,(\omega/v_F q)$, where $N_f=4$ accounts for spin-valley degeneracy, $\varepsilon_F$ is the Fermi energy, $v_F$ the Fermi velocity, $q$ the momentum, and $\omega$ the frequency. Comparing with Eq.~(\ref{eq:chiT_expand}), we identify $\alpha = (N_f \varepsilon_F)/(2\pi\hbar^2)$. As $\varepsilon_F = 0.1\,{\rm eV}$ for graphene and  $\varepsilon_F = 1\,{\rm meV}$ for MATBG, this yields $\alpha = 9.2 \times 10^{47}\ {\rm J^{-1}s^{-2}}$ and $\alpha = 9.2 \times 10^{45}\ {\rm J^{-1}s^{-2}}$ respectively.

We now estimate $\bar{\chi}_{\rm orb}$ and $\bar{\alpha}$ using the definitions, $\bar\chi_{\rm orb}=g_2^2 k_{\rm F}^2 \chi_{\rm orb}/(v_{\rm F}^4\rho_m)$ and
$\bar\alpha=(g_2^2\alpha)/(v_{\rm F}^4\rho_m)$. For graphene, we have $v_{\rm F}=10^{6}~\rm {m~s^{-1}}$, $\varepsilon_{\rm F}=0.1~ \rm eV$ and  find that \hypertarget{chibarorb}$\bar \chi_{\rm orb} = -1.72 \times10^{-8}$ and $\bar{\alpha} = 7 \times 10^{-8}$. In the case of MATBG, we have $v_{\rm F}=10^{4}~\rm {m ~s^{-1}}$ and $\varepsilon_{F} = 1~\rm meV$ find that $\bar \chi_{\rm orb} = -1.72 $ and $\bar{\alpha} = 7\times 10^{-2}$.

To estimate a representative paramagnetic value of the orbital susceptibility, we use the results of Ref.~\cite{principiparamagnetisim}, where the orbital magnetic response of doped graphene is found to be finite and positive due to electron--electron interactions. For a carrier density $n\sim10^{12}\,{\rm cm^{-2}}$ (consistent with the $k_{F}$ used), the orbital susceptibility per unit mass is reported to be $\chi_{\rm mass}= 4\times10^{-6}\,{\rm emu/g}$  \cite{principiparamagnetisim}. Using $1\,{\rm emu/g}=4\pi\times10^{-3}\,{\rm m^3\,kg^{-1}}$, this corresponds to a mass susceptibility $\chi_{\rm mass}= 4\times10^{-6}\times 4\pi\times10^{-3}= 5.0\times10^{-8}\,{\rm m^3\,kg^{-1}}$. Multiplying by the areal mass density of graphene, $\rho_m=7.6\times10^{-7}\,{\rm kg\,m^{-2}}$, we obtain the corresponding two-dimensional magnetic susceptibility $\chi_m= \rho_m \chi_{\rm mass}= 3.8\times10^{-14}\,{\rm m}$. Finally, using the relation $\chi_m=(\mu_0 e^2/4\pi)\,\chi_{\rm orb}$, we obtain $\chi_{\rm orb}= \chi_m/(\mu_0 e^2/4\pi)= 1.5\times10^{31}\,{\rm m}^2\,{\rm J}^{-1}\,{\rm s}^{-2}$. This is of the same order of magnitude as the diamagnetic lattice estimate used in the main text, but with opposite sign. Using the definitions of the dimensionless parameters, this corresponds to $\bar{\chi}_{\rm orb} = 2.6\times10^{-8}$ for graphene and $\bar{\chi}_{\rm orb}= 2.6$ for MATBG.
\clearpage
\bibliography{main.bib}
\end{document}